\documentclass[aps,prb,twocolumn,showpacs,superscriptaddress,nofootinbib,longbibliography]{revtex4-1}

\usepackage{latexsym}
\usepackage{amssymb}
\usepackage{graphicx}
\usepackage{amsmath}
\usepackage{tikz}
\usepackage{bm}
\usepackage[colorlinks,
linkcolor=black,
citecolor=black,
urlcolor=blue
]{hyperref}
\usepackage{verbatim}
\usepackage{mathrsfs}
\usepackage{extarrows}
\usepackage{comment}
\usepackage{mathtools,slashed}
\usepackage{soul}
\usepackage[toc,page]{appendix}
\usepackage[vcentermath]{youngtab}
\usepackage{multirow}
\usepackage[width=.75\textwidth]{caption}
\usepackage{atbegshi,picture}
\usepackage{lipsum}
\DeclareMathOperator{\tr}{tr}

\makeatletter
\renewcommand\@makecaption[2]{
	\par
	\vskip\abovecaptionskip
	\begingroup
	\small\rmfamily
	\begingroup
	\samepage
	\flushing
	\let\footnote\@footnotemark@gobble
	\@make@capt@title{#1}{#2}\par
	\endgroup
	\endgroup
	\vskip\belowcaptionskip
}
\makeatother

\AtBeginShipoutNext{\AtBeginShipoutUpperLeft{
		\put(\dimexpr\paperwidth-1cm\relax,-1.5cm){\makebox[0pt][r]}
}}

\makeatletter\begin{document}

	\title{Boundary conditions and anomalies of conformal field theories in 1+1 dimensions}
\author{Linhao Li}   
\affiliation{Department of Physics and Astronomy, University of Ghent, 9000 Ghent, Belgium}
\affiliation{Institute for Solid State Physics, The University of Tokyo. Kashiwa, Chiba 277-8581, Japan}
\author{Chang-Tse Hsieh}         
\thanks{cthsieh@phys.ntu.edu.tw}
\affiliation{Department of Physics and Center for Theoretical Physics, National Taiwan University,
Taipei 10607, Taiwan}
\affiliation{Physics Division, National Center for Theoretical Science, National Taiwan University,
Taipei 10607, Taiwan}

\author{Yuan Yao}
\thanks{smartyao@sjtu.edu.cn}
\affiliation{School of Physics and Astronomy, Shanghai Jiao Tong University, Shanghai 200240, China}	
	
		\author{Masaki Oshikawa} \affiliation{Institute for Solid State Physics, The University of Tokyo. Kashiwa, Chiba 277-8581, Japan}        
\affiliation{Kavli Institute for the Physics and Mathematics of the Universe (WPI),The University of Tokyo, Kashiwa, Chiba 277-8583, Japan}
\affiliation{Trans-scale Quantum Science Institute, University of Tokyo, Bunkyo-ku, Tokyo 113-0033, Japan}
	\date{\today}

	\begin{abstract}

We study a relationship between conformally invariant boundary conditions and anomalies of conformal field theories (CFTs) in 1+1 dimensions. For a given CFT with a global symmetry, we consider symmetric ``gapping potentials'' which are relevant perturbations to the CFT. If a gapping potential is introduced only in a subregion of the system, it provides a certain boundary condition to the CFT.
From this equivalence, if there exists a Cardy boundary state which is invariant under a symmetry,
then the CFT can be gapped with a unique ground state by adding the corresponding gapping potential.
This means that the symmetry of the CFT is anomaly free.
Using this approach, we systematically deduce the anomaly-free conditions for various types of CFTs with several different symmetries.
They include the free compact boson theory, Wess-Zumino-Witten models, and unitary minimal models.
When the symmetry of the CFT is anomalous, it implies a Lieb-Schultz-Mattis type ingappability of the system.
Our results are consistent with, where available, known results in the literature.
Moreover, we extend the discussion to other symmetries including spin groups and generalized time-reversal symmetries.

	\end{abstract}
	
	\maketitle

	\section{Introduction}
	Symmetry often plays an essential role in the study of quantum many-body physics. 
	In classical physics, it implies various conservation laws and corresponding conserved currents. However, on the quantum level, there is a potential obstruction in promoting a global symmetry to a gauge symmetry which is called 't Hooft anomaly~\cite{hooft1980naturalness}. Recent works on topological phases of matter have revealed that a $d$-dimensional quantum field theory with a global symmetry having a 't Hooft anomaly can be regarded as the boundary theory of a nontrivial symmetry protected topological (SPT) phases \cite{gu2009tensor,chen2011classification,chen2011complete,hasan2010colloquium,qi2011topological,fu2011topological} in one higher dimension and the anomaly of the boundary theory corresponds to certain topological property of the bulk theory~\cite{chen2013symmetry,chen2012symmetry,PhysRevB.85.075125,fidkowski2011topological,PhysRevB.86.125119}.

	For a (1+1)d conformal field theory (CFT) with an anomalous global symmetry $G$, the anomaly can be detected by examining modular invariance of the partition function(s) of the CFT orbifolded by $G$ \cite{PhysRevB.85.245132, PhysRevB.88.075125}. 
(In a $G$-orbifold CFT, states except for $G$ singlets are projected out and $G$-twisted sectors are included when evaluating the partition function.)
If $G$ has a 't Hooft anomaly, it is impossible to construct a modular invariant $G$-orbifold partition function from the original CFT.

	Another probe of anomaly is edgeability \cite{vishwanath2013physics,chen2016bulk,thorngren2021anomalous}. A CFT with a global symmetry $G$ is edgeable means it can be cut open into a boundary conformal field theory (BCFT) where $G$ is preserved both in the bulk and on the boundary. The connection between edgeability and anomaly relies on the observation that a (1+1)d CFT with an anomalous symmetry lives on the boundary of a (2+1)d nontrivial SPT phase and thus can not be formulated consistently on a one-dimensional space with a boundary (as the boundary of a boundary vanishes)~\cite{han2017boundary,wang2013non,wang2019solution,Choi:2023xjw,Chen:2023hmm,Thorngren:2020yht,Wang:2024vjl}. That is, a (1+1)d CFT with an anomalous symmetry is not edgeable in a symmetry-preserving manner to a BCFT.

	In this work, we further investigate the correspondence between boundary conditions  and anomalies of CFTs from a different aspect. We consider adding spatially dependent relevant perturbations to a CFT such that the perturbations are present only outside a finite length segment of the one-dimensional space $\mathbb{R}^1$.  The boundary conditions at the interfaces between the regions with and without perturbations are determined by the forms of the added relevant operators and undergo renormalization group (RG) flow with fixed points corresponding to conformal boundary conditions. It turns out that the theory on the finite segment will become a BCFT with these conformal boundary conditions\cite{cho2017relationship}.

	We will focus on the uniqueness of ground states after adding particular space-dependent perturbation. More precisely, in the limit length $L\to 0$, the multiplicity of the identity operator contained in the partition function equals ground-state degeneracy. In order to have a unique ground state, the identity operator should appear in the partition function just once which means the partition function goes to 1 in this limit. We will show this requirement is equivalent to finding a symmetric boundary state or boundary condition. 
	On the contrary, if all the possible boundary amplitudes (partition function) with an imposed symmetry exhibit a degeneracy, it suggests that the CFT cannot be gapped with a unique ground state, i.e., ingappability. The latter property implies an anomaly of the CFT. 

If we use the gapping-potential argument directly to deduce whether a symmetry is anomaly-free, we need to first construct a symmetric gapping potential and then check whether the gapped ground state(s) does not break the symmetry spontaneously.
This procedure can be systematically implemented on free compact boson CFTs but is usually not easy for most other CFTs, such as WZW models and minimal models. On the other hand, the BCFT formulation and the construction of certain boundary states of several CFTs, including the two mentioned above, have been known to us, so we can check the existence of symmetric boundary states systematically and quickly using the BCFT approach. For example, for the   $\mathrm{SU}(2)_{1}$ WZW  model  with  the  T-duality symmetry, we will show the boundary state approach can give the symmetric gapping potentials which are beyond the usual Haldane null gapping potentials~\cite{PhysRevLett.74.2090}. Moreover, the boundary states approach can also imply the anomaly with respect to the time reversal symmetry of WZW models, which are usually more difficult to be detected by other approaches.

An analogous application of BCFT was discussed in Ref.~\onlinecite{cho2017relationship}.
There, the SPT phases in $1+1$ dimensions were related to the spectrum of the ``mother CFT'' with distinct boundary
conditions corresponding to the trivial and SPT phases imposed at the two ends.
In the present paper, we rather discuss the anomaly of the CFT, which is related to the
Lieb-Schultz-Mattis (LSM) type ``ingappability''~\cite{LIEB1961407, Affleck:1986aa, Oshikawa2000Commensurability, PhysRevB.69.104431,PhysRevB.83.035107,PhysRevB.93.104425,ogata2019lieb,ogata2021general,PhysRevLett.126.217201,Tasaki:2022gka,Li:2022nwa,Cheng:2022sgb,Seifnashri:2023dpa}
, by imposing the ``same'' boundary conditions at the two ends.
Thus our discussion is complementary to Ref.~\onlinecite{cho2017relationship}.
(For previous works on anomaly-based arguments for the LSM type ingappability, see Ref.~\onlinecite{Cheng:2016aa, Shunsuke-Oshikawa17, Cho:2017aa, Metlitski-Thorngren17, Tanizaki2018, Yao-Hsieh-Oshikawa19,Cheng:2022sgb,Seifnashri:2023dpa})

Furthermore, a similar idea was used to discuss the anomaly of WZW models between the center symmetry and the spacetime (large) diffeomorphism~\cite{numasawa2018mixed}.
While this was useful in revealing the boundary ingappability of SPT phases,
its application on LSM-type ingappability --- ingappability induced from some internal symmetry and lattice symmetry ---
was still unclear.
In contrast, in our work,
we provide a classification of mixed anomalies of
e.g., a large class of Lie groups and their centers,
where the Lie groups can be the internal symmetry and the center symmetry is expected to realize the lattice symmetry.
Thus, our results are more directly relevant to LSM-type ingappability of lattice models.
Indeed, our results are consistent with known ingappabilities found on lattices.
Moreover, our argument covers a broad class of 1d spin systems. One class of such systems that has been rarely studied is the $\mathrm{SO}(n)$ spin chains, whose associated LSM theorem was discussed in Ref.~\onlinecite{Tu2011EffectiveFT, Jian-Bi-Xu18} and is consistent with the results presented here.

The organization of the rest of the paper is as follows.
	In section \ref{2}, a brief introduction to BCFT is provided. In section \ref{sec 3}, we will discuss the correspondence between boundary conditions and anomalies of CFTs in 1+1 dimensions. 
	As an application and illustration of our framework, we study some concrete examples. In section \ref{sec 4}, the compact boson CFTs in 1+1 dimensions with $\mathrm{PSU}(N)$ and  $\mathbb{Z}_{N}$ symmetry is considered. In section \ref{sec 5}, we will apply our framework to anomalies of  WZW models with center symmetry, vector rotation symmetry and time reversal symmetry. In section \ref{sec 7}, we will show discrete global symmetries of minimal models are all anomaly-free.   In section \ref{sec 6}, we will discuss the anomaly of free boson theory with respect to the T-duality symmetry. Our main results are summarized in TABLE \ref{table4}.

\begin{table*}[t]
		\centering
		\begin{tabular}{c c c c c c }
			\hline\hline
			Model & Symmetry &Invariant boundary states \\
			\hline		
		        $\mathrm{SU}(N)_{k}$ \text{WZW model}&&$k\in$$N$$\mathbb{N}$\\
			\\[-1em]
			$\mathrm{Spin}(2N+1)_{k}$ \text{WZW model}&&$k\in$$\mathbb{N}$\\
			\\[-0.7em]
			$\mathrm{Usp}(N)_{k}$ \text{WZW model} &$\text{diagonal rotation symmetry}$&$k\in$$2\mathbb{N}$ or $N\in$$2\mathbb{N}$\\
			\\[-0.9em]
			$\mathrm{Spin}(4N+2)_{k}$ \text{WZW model}&\&&$k\in$4$\mathbb{N}$ \\
			\\[-0.7em]
			$\mathrm{Spin}(4N)_{k}$ \text{WZW model}&center symmetry&$k\in$2$\mathbb{N}$\\
			\\[-0.7em]
			
			$E_{6}$ \text{WZW model}&&$k\in$$3\mathbb{N}$\\
			\\[-0.7em]
			$E_{7}$ \text{WZW model}&&$k\in$$2\mathbb{N}$\\			\\[-1em]
			\hline

			$\mathrm{SU}(N)_{k}$ \text{WZW model}&&$k\in$$2\mathbb{N}$ or $N\in$$4\mathbb{N}$\\
			\\[-1em]
			$\mathrm{Spin}(2N+1)_{k}$ \text{WZW model}&&$k\in$$\mathbb{N}$\\
			\\[-0.7em]
			$\mathrm{Usp}(N)_{k}$ \text{WZW model} &$\text{diagonal rotation symmetry}$&$k\in$$2\mathbb{N}$ or $N\in$$2\mathbb{N}$\\
			\\[-0.9em]
			$\mathrm{Spin}(4N+2)_{k}$ \text{WZW model} &\&&$k\in$$\mathbb{N}$ \\
			\\[-0.7em]
			$\mathrm{Spin}(4N)_{k}$ \text{WZW model}&time-reversal symmetry $\mathcal{T}_2$&$k\in$2$\mathbb{N}$\\
			\\[-0.7em]

			$E_{7}$ \text{WZW model}&&$k\in$$2\mathbb{N}$\\			\\[-1em]
			\hline
			 $\mathrm{SU}(2)_{1}$ \text{WZW model}&T-duality&yes\\
			\\[-1em]
			
				(self-dual compact boson)&T-duality extended by center symmetry&no\\
			\\[-1em]
			\hline
			 \text{unitary (Virasoro) minimal models}\quad& any finite symmetry i.e. $\mathbb{Z}_2$ or $S_3$&yes\\
			\\[-1em]
			\hline\hline
		\end{tabular}
\caption{Summary of the conditions for the existence of symmetry invariant boundary states in various CFTs. Here we denote the $G_k$-WZW models at the level $k$ by the associated Lie groups $G$.}\label{table4}
	\end{table*}

	\section{A review on Boundary Conformal Field Theory}\label{2}
	In this section, we will give a basic review of the BCFT on the upper half plane and annulus,
which is an important basis for our formulation.

	\subsection{BCFT on the Upper Half Plane}
	On the complex plane, the analytic coordinate transformations form the conformal group. If we put the CFT on the Upper Half Plane (UHP), the conformal transformation should also map UHP to itself \cite{cardy2004boundary,kawai2002boundary,recknagel2013boundary}:
	\begin{eqnarray}
	(x,y)\to(x,y)+(\epsilon_x,\epsilon_y)\,\,\,\text{and } \epsilon_{y}(x,0)=0.
	\end{eqnarray}
	
	To see the effect of boundary conditions, 
	we can calculate the expectation value of changes of primary fields $X$ transformed by the above conformal transformation:
	\begin{eqnarray}
	&&\int d\Phi \delta X e^{-S[\Phi]}
	=\int d\Phi X \delta S e^{-S[\Phi]}\nonumber\\&&=-\int dx \epsilon_{x}(x,0)\langle T^{xy}(x,0)X\rangle\nonumber\\&&+\int dxdy \epsilon_{\mu}\partial_{\nu}\langle T^{\mu\nu}(x,y)X\rangle.
	~
	\end{eqnarray}
	The second term generates the conformal transformation on $X$ producing $\langle\delta X\rangle$, which demands the first term to vanish, i.e., $T^{xy}(x, 0) = 0$. This condition can be interpreted as the absence of energy flow across the boundary.
	
	In the complex coordinate, this boundary condition can be written as:
	\begin{eqnarray}
	T(z)=\bar{T}(\bar{z}),\quad z\in \mathbb{R}.
	\end{eqnarray}
	Thus from the definition of the generators of the conformal group expanding the Laurent series of energy-momentum tensors, we obtain
	\begin{eqnarray}
	T(x)=L_{n}x^{-n-2}=\bar{T}(x)=\bar{L}_{n}x^{-n-2}\Rightarrow L_{n}=\bar{L}_{n}.
	\end{eqnarray}
 If we see the $y$-axis as the time direction, then the Hilbert space is defined on the $x$-axis.
	Thus, in contrast to the case of CFTs on the full complex plane, only one copy of Virasoro algebra acts on the Hilbert space $
	\oplus n_{h} V_{h}$. 
	
	\subsection{BCFT on the annulus}
	We can also define CFT on an annulus. The boundary can be placed perpendicular to the spatial direction (open string) or to the time direction (closed string but with time open). 
	
	In the closed string picture, the boundary state is represented by a state in the Hilbert space of a CFT defined on a circle and boundary A is on the time $t=0$ and boundary B is on $t=L$. 
	For spatial dimension, we identify 
	$x+it\sim x+\beta+it$. Then we let $\omega=x+it$ and map the cylinder into the plane: $z=\exp({-i 2\pi}\omega/{\beta})$. The boundary is mapped to the circle          $|z|=1$ and $|z|=\exp({2\pi L}/{\beta})$. From the calculation in the above section, we can see the conformal boundary conditions  should be: $(c=\bar{c})$
	\begin{eqnarray}
	z^2T(z)-c/24=\bar{z}^2\bar{T}(\bar{z})-\bar{c}/24\ , \quad|z|=1,\exp({2\pi L}/{\beta}).\nonumber\\
	\end{eqnarray}

	More precisely,  after quantization, the above equation satisfies when it acts on the boundary state in the Heisenberg picture.  
	Using the mode expansion of the stress-energy tensor, we can obtain
	\begin{eqnarray}
	(L_{n}-\bar{L}_{-n})|a\rangle\rangle=0.
	\end{eqnarray}
	
	Thus the physical boundary condition that we put on the boundary should be consistent with the conformal boundary conditions. The physical  boundary condition describes an automorphism from the holomorphic sector to the antiholomorphic sector
	\begin{eqnarray}\label{conformal boundary}
	\lbrack S(z)-\rho_{\text{A/B}}({\bar{S}(\bar{z})})\rbrack|a\rangle\rangle=0,
	\end{eqnarray} 
	where $S$ belongs to some symmetry algebra and $\rho_{\text{A/B}}$ denotes an automorphism of the algebra of fields.
	For example, the U(1) symmetry in the compact boson CFT has the current $i\partial \varphi$ and the boundary condition can be: $\partial_{z} \varphi_{z}=\pm \partial_{\bar{z}} \varphi_{\bar{z}}|_{|z|=1,\exp(2\pi L/\beta)} $.
	
	A kind of   quantum   states  $|a\rangle\rangle$ satisfying the equation ($\ref{conformal boundary}$) is called Ishibashi  states\cite{Ishibashi1989THE}. They have the following properties:
	\begin{eqnarray}
	\langle\langle b| e^{-2\pi \frac{L}{\beta}(\hat{H}_{L}+\hat{H}_{R})}|a\rangle\rangle=\delta_{ba}\chi_{a}(\tilde{q}=e^{-4\pi \frac{L}{\beta}}),
	\end{eqnarray}
 where $\hat{H}_{L}=L_0-c/12$ and $\hat{H}_{R}=\bar{L}_0-c/12$ are  the Hamiltonians of the holomorphic and antiholomorphic degrees of freedom.
	Here $\chi_{a}$ is the character of an irreducible representation of the CFT.
	
	Then the general boundary states can be expanded in terms of Ishibashi states:
	\begin{eqnarray}
	|A\rangle=A_{a}|a\rangle\rangle.
	\end{eqnarray}
	
	The partition function can be written as an amplitude of boundary states:
	\begin{eqnarray}
	Z_{\text{AB}}=\langle\Theta A| e^{-2\pi \frac{L}{\beta}(\hat{H}_{L}+\hat{H}_{R})}|B\rangle.
	\end{eqnarray}
	Now $\beta$ is the circumference of space direction with periodic boundary conditions (a circle) and $L$ is propagating time.
	Here $\Theta$ is a CPT operator since these two boundaries have opposite orientations:
	\begin{eqnarray}
	\Theta c \Theta^{-1}=c^{*}, \quad\Theta|B\rangle=(B_{a})^{*}|a^{+}\rangle\rangle,
	\end{eqnarray}
	where $|a^{+}\rangle\rangle=C|a\rangle\rangle$ and $C$ is the charge conjugation operator. In most CFTs, the CPT operator $\Theta$ acts as the identity operator.
	
	Then, the partition function can be written as
	\begin{eqnarray}
	Z_\text{AB}=\sum_{a}A_{a}B_{a}\chi_{a}(\tilde{q}=e^{-4\pi \frac{L}{\beta}}).
	\end{eqnarray}
	
	In the open string picture, BCFT is a CFT with boundary conditions specified by A($x=0$) and B($x=L$) at these two conformal boundaries in spatial direction and 
	$t+ix\sim t+\beta+ix$ identified in the time direction. Since these two pictures can be related by the  S-modular transformation --- a $90$-degree rotation of the space-time manifold, the symmetry algebra will transform in the following way:
	\begin{eqnarray}
	\ln z=i\ln z'\ , S(z)=i^{h}S(z')\frac{z'}{z}, \bar{S}(\bar{z})=i^{-\bar{h}}S(\bar{z}')\frac{\bar{z}'}{\bar{z}},\nonumber\\
	\end{eqnarray}
 where $h$ and $\bar{h}$ are the conformal weights of $S$ and $\bar{S}$.
	Then the boundary condition in the open chain is obtained from that of closed string picture under the S-modular transformation:
	\begin{eqnarray}
	S(z)=\rho_{\text{AB}}({\bar{S}(\bar{z})})\longrightarrow S(z')=(-1)^{h}\rho_{\text{AB}}({\bar{S}(\bar{z}')}),\nonumber\\
	\end{eqnarray}
	where $|z'|=1,\exp({2\pi L}/{\beta}).$
	In this picture, the boundary condition constrains the Hibert space of quantum states on the interval which is denoted by $\mathcal{H}_\text{AB}$. The partition function (at inverse temperature $\beta$ ) is written as a trace of the Hamiltonian $\hat{H}^\text{open}_\text{AB}$ of the finite interval of length  with boundary conditions A and B :
	\begin{eqnarray}
	Z_\text{AB}=\tr_{\mathcal{H}_\text{AB}} e^{-\beta \hat{H}^\text{open}_\text{AB}}.
	\end{eqnarray}
	
	The partition function can be rewritten using the Hamiltonian defined purely in the holomorphic sector since only holomorphic (left-moving) degrees of freedom contribute:
	\begin{eqnarray}
	Z_\text{AB}=\tr_{\mathcal{H}_\text{AB}} q^{\hat{H}^{L}}.
	\end{eqnarray}
	All terms in the partition function are powers of
	$
	q=e^{-\pi{\beta}/{L}}
	$
	related to the length $L$ and inverse temperature $\beta$.
	
	Since only holomorphic (left-moving) degrees of freedom appear in the BCFT, the partition function can be decomposed into characters of different irreducible representations $\phi_{a}$ of the holomorphic Virasoro algebra  \cite{cardy1989boundary}:
	\begin{eqnarray}
	Z_\text{AB}=\sum_{a}n^{a}_\text{AB}\chi_{a}(q).
	\end{eqnarray}
	The non-negative integers $n^{a}_\text{AB}$ represent the multiplicity with which the irreducible representations appear in the Hibert space $\mathcal{H}_\text{AB}$.
	
	These characters in the open string and closed string  are related by a modular transformation of the space-time torus:
	\begin{eqnarray}
	\chi_{a}(q)=\sum_{b} S_{ab}\chi_{b}(\tilde{q}).
	\end{eqnarray}
	Now the Cardy condition requires that this expression in the closed picture can be interpreted as that in the open picture. That is, the non-negative integers and the expansion coefficients can be related via \cite{cardy1989boundary}:
	\begin{eqnarray}
	n^{a}_{\text{AB}}=\sum_{b} A_{b}B_{b}S_{ab}.
	\end{eqnarray}

	\section{Boundary Conformal Field Theories
		and quantum anomalies}\label{sec 3}
	
	In this section, we give a set of arguments which support the advocated relation between BCFTs and gappabilities of CFTs.
	
	We start with
	a brief overview of gappabilities of CFTs. A given CFT in (1+1) dimensions can be gapped by a ``massive deformation'' of a
	CFT,
	\begin{eqnarray}
	S\to S-\lambda\int dtdxV(\phi(x)),
	\end{eqnarray}
	where $V(\phi(x))$ is a relevant operator, and $\lambda\in\mathbb{R}$ is the coupling constant.
	
	To construct boundary conditions, one can also consider a relevant operator which depends on the space coordinate. More precisely, one can consider the following relevant operator:
	\begin{eqnarray}
	V(x)=\begin{cases}
	
	V_\text{A}(\phi(x)),&x<0;\\
	
	0,&0 \leq x < L;\\
	V_\text{B}(\phi(x)),&L \le x.
	
	\end{cases}
	\label{eq:2bc}
	\end{eqnarray}

	If both $V_\text{A}(\phi)$ and $V_\text{B}(\phi)$  can gap the system with a unique ground state, only the middle region is still gapless in the sense that the bulk gapless modes have nonzero amplitude only there. At low energy, the boundaries between the CFT and any of the neighborhood gapped phases are expected to be renormalized into conformally invariant boundary conditions\cite{cho2017relationship}. In other words, this implies that there is a relationship between relevant operators and conformal invariant boundary conditions.

	For our interest, the interactions are assumed to be related by a symmetry $G$.
	That is, in the setup~\eqref{eq:2bc}, we choose
	\begin{equation}
	    V_\text{B} = 	gV_\text{A}(\phi(x))g^{-1} ,
	    \label{symmV}
	\end{equation}
	where $g$ is an element of the symmetry group $G$, as shown in Fig.~\ref{fig:V_setup}.
	~\\~\\~\\
	\begin{frame}
		
		\begin{tikzpicture}  
		\draw[->](-3,0)to(5,0); 
		\node at(5.2,0) {$x$};
		\draw (0,0) -- (0,1)[dashed];
		\draw (2,0) -- (2,1)[dashed];
		\node at(-1.5,0.3) {$V=V_{\text{A}}(\phi)$};
		\node at(1,0.3) {$V=0$};
		\node at(3.5,0.3) {$V=gV_{\text{A}}(\phi)g^{-1}$};
		\node at(0,-0.3) {$x=0$};
		\node at(2,-0.3) {$x=L$};
		\end{tikzpicture}
		
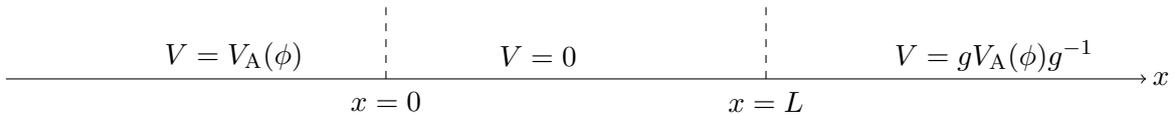
\captionof{figure}{\textbf{space-dependent interaction with respect to transformation $g$.}}
		\label{fig:V_setup}
	\end{frame}

%
%
%
%
%
%

	The corresponding boundary conditions should also be related via the same transformation. Now if $V_{\text{A}}(\phi)$ is symmetric, the interactions in the region $x<0$ and $x>L$ are the same. In the limit $L\to0$ ($q\to 0$), the theory is fully gapped and the partition function of BCFT is just that of this gapped phase which is 1:
	\begin{eqnarray}
	Z_\text{AB}(q)\to 1,\quad L\to 0.\label{con:anomaly}
	\end{eqnarray}
	If we can find a boundary condition A satisfying Eq.~(\ref{con:anomaly}) for every $g\in G$, we can argue that the symmetry $G$ is anomaly-free according to the gapping potential argument.
	
	On the other hand, in the closed string picture, the length $L$ plays a role of time. Thus the equation ~(\ref{con:anomaly})  implies the boundary state $|A\rangle$ is invariant under each symmetry transformation $g$:
	\begin{eqnarray}\label{bs}
	Z_\text{AB}(L\to 0)=\langle\Theta A|g|A\rangle=\langle A|g|A\rangle=1,~ \forall g\in G.
	\end{eqnarray}
	Here we use the fact that the CPT operator acts as the identity operator for most CFTs.  Therefore the boundary condition A and interaction $V_{\text{A}}(\phi)$ should be also symmetric.
	
	In fact, from an argument about the close relationship between boundary state and ground state after adding gapping potential, the smeared boundary state $e^{-\tau H_\text{CFT}}|A\rangle $ can serve as the possible ground state after adding the relevant operator $V_\text{A}(\phi)$ \cite{SciPostPhys.3.2.011}. Since  Hamiltonian of the CFT commutes with symmetry $G$, this possible ground state is symmetric if and only if the boundary state is symmetric. 
	
	For example, the lsing CFT has a $\mathbb{Z}_{2}$ symmetry which flips every spin and three boundary states $|\uparrow\rangle$, $|\downarrow\rangle$, $|\text{free}\rangle$. Since the boundary state $|\text{free}\rangle$ is symmetric, the corresponding smeared boundary state is a symmetric ground state and the corresponding perturbation is $V=t\int dx\ \hat{\epsilon}(x),\ t>0$ where $\hat{\epsilon}$ is energy density \cite{SciPostPhys.3.2.011}.

	\section{ Multicomponent U(1) free compact boson CFTs}\label{sec 4}
	In this section, we will use the boundary state approach to calculate the anomaly of the free boson CFTs with respect to some global symmetries.
	\subsection{Canonical quantization}
	Multicomponent free boson theories are a general class of CFTs which can be realized at the edge of (2+1) dimensional gapped many-body systems belonging to the SPT phase \cite{PhysRevB.86.125119}:
	\begin{eqnarray}
	S=\frac{1}{8\pi }\int dxdt[K_{IJ}\partial_{t}\phi^{I}\partial_{x}\phi^{J}-V_{IJ}\partial_{x}\phi^{I}\partial_{x}\phi^{J}],
	\end{eqnarray}
	where $K$ matrix is $\oplus^{N}_{i=1}\sigma^{x}$ and $I$, $J$ = 1, $\cdots$ , $2N$. This theory has $N$ copies of non-chiral bosons, $V$ is a $2N\times 2N$ symmetric, positive definite real matrix and contains the information like the velocities of the edge modes. Here we choose the $V$ matrix to be an identity matrix in this paper since it does not influence the topological properties of this theory. If this CFT is realized on the  cylinder of circumference $L$, the compact U(1) boson fields $\phi^{I}$ should satisfy the compact condition with radius $R=\sqrt{2}$:
	\begin{eqnarray}
	\phi^{I}(x+L,t)= \phi^{I}(x,t)+2\pi n^{I}\sqrt{2}, \quad n^{I}\in\mathbb{Z}.
	\end{eqnarray} 
	We can also write the action above in the usual way:
	\begin{eqnarray}
	S=\frac{1}{8\pi}\int dxdt \sum^{I=N}_{I=1} \partial_{\mu}\phi^{2I-1}\partial^{\mu}\phi^{2I-1}.
	\end{eqnarray}
	We define variable $\phi^{2I}$ as the dual variable $\theta^{I}$.
	
	It is more convenient to carry out the quantization in the chiral basis $\varphi_{I}$ which we define by diagonalizing the $K$ matrix as:
	\begin{eqnarray}
	P\phi=\varphi,\quad P K P^{T}=2\eta,
	\end{eqnarray} 
	where $P$ is $2N\times 2N$ orthogonal matrix and $\eta$ is a diagonal matrix with $\eta_{ii}=(-1)^{i-1}$. 
	
	The action can be rewritten in the chiral basis as follows:
	\begin{eqnarray}
	S&&=\frac{1}{4\pi}\int dx^{2}[\partial_{t}\varphi^{I}_{L}{x}\varphi^{I}_{L}-\partial_{t}\varphi^{I}_{R}{x}\varphi^{I}_{R}\nonumber\\&&-\partial_{x}\varphi^{I}_{L}\partial_{x}\varphi^{I}_{L}-\partial_{x}\varphi^{I}_{R}\partial_{x}\varphi^{I}_{R}].
	\end{eqnarray}
	In this picture, chiral bosons satisfy the compactified condition:
	\begin{eqnarray}
	&&\varphi^{I}_{L}(x+L,t)= \varphi^{I}_{L}(x,t)+2\pi n^{I}R, \quad n^{I}\in\mathbb{Z},\nonumber\\
	&&\varphi^{I}_{R}(x+L,t)= \varphi^{I}_{R}(x,t)+2\pi m^{I}R, \quad m^{I}\in\mathbb{Z}.
	\end{eqnarray}
	After canonical quantization, the boson fields satisfy the commutation relations:
	\begin{eqnarray}
	&&[\varphi^{I}_{L}(x),\partial_{x}\varphi^{J}_{L}(x')]=2\pi i\delta^{IJ}\delta(x-x'),\nonumber\\
	&&\lbrack\varphi^{I}_{R}(x),\partial_{x}\varphi^{J}_{R}(x')\rbrack=-2\pi i\delta^{IJ}\delta(x-x').
	\end{eqnarray}
	The equation of motion is 
	\begin{eqnarray}
	&\partial_{t}\varphi^{I}_{L}-\partial_{x}\varphi^{I}_{L}=0,\nonumber\\
	&\partial_{t}\varphi^{I}_{R}+\partial_{x}\varphi^{I}_{R}=0.
	\end{eqnarray}
	As a result, the mode expansion is 
	\begin{eqnarray}
	\varphi^{I}(x,t)_{L/R}=\varphi_{0}+\frac{2\pi}{L}(t\pm x)a^{I}_{0}+\sum_{r\ne 0}a^{I}_{r}e^{-\frac{2\pi ri}{L}(t\pm x)}.\nonumber\\
	\end{eqnarray}
	The canonical commutation relations and the compactified conditions for the operators are
	\begin{eqnarray}
	&&[a^{I}_{r,L/R},a^{J}_{s,L/R}]=r\delta^{IJ}\delta_{r+s,0},\nonumber\\
	&&\varphi^{I}_{0,L/R}\sim \varphi^{I}_{0,L/R}+2\pi n^{I}R,\nonumber\\
	&&\lbrack \varphi^{I}_{0,L/R},a^{J}_{0,L/R} \rbrack=\pm  i \delta^{IJ}.
	\end{eqnarray}
	And the eigenvalue $a^{I}_{0}$ is quantized to be $m^{I}$, where $\sqrt{2}m^{I}\in \mathbb{Z}$.
	\subsection{ Anomaly free condition and Haldane gapping potential}
	In the compact boson CFTs, the conformal boundary conditions for Ishibashi states are:
	\begin{eqnarray}
	(L_{r}-\bar{L}_{-r})|K\rangle\rangle=0.
	\end{eqnarray}
	Here $L$ and $\bar{L}$ are Virasoro generators in each chiral sector. A special solution for this equation is given by \cite{oshikawa2010boundary}:
	\begin{eqnarray}
	(a^{I}_{r,L}-D_{IJ}a^{J}_{-r,R})|v\rangle\rangle=0.
	\end{eqnarray}
	The Ishibashi states for this condition are:
	\begin{eqnarray}
	|v\rangle\rangle=\exp\left(\sum_{r=1}^{\infty}\frac{1}{r}a^{I}_{-r,,L}D_{IJ}a^{J}_{-r,,R}\right)|v\rangle,
	\end{eqnarray}
	where $|v\rangle$ are eigenstates of $a^{I}_{0,L}$. And its eigenvalues $v^{I}$ satisfy $\sqrt{2}v^{I}\in \mathbb{Z}$. 	This eigenstate can be written as a coherent state,
	namely, 
	\begin{eqnarray}
	|v\rangle=e^{i\sum_{I}v^{I}\lbrack\varphi^{I}_{0,L}+(D\varphi_{0,R})^{I})\rbrack}|0\rangle .
	\end{eqnarray}
	The matrix $D$ can be any $N\times N$ orthogonal matrix with eigenvalue $d^{I}=$ 1 or $-$1. The eigenvector is denoted as $e_{I}$.  
	
	The Cardy states can be constructed as the superposition of Ishibashi states:
	\begin{eqnarray}
	|B,\lbrace\alpha^{I}\rbrace\rangle=\otimes_{I} |B,\alpha^{I}\rangle,\ 
	|B,\alpha^{I}\rangle=\sum_{v^{I}} e^{iv^{I}\alpha^{I}}	|v^{I}\rangle\rangle.	
	\end{eqnarray}
	Such Cardy states can be rewritten as follows:
	\begin{eqnarray}
	|B,{\alpha_{j}}\rangle=&& e^{\sum_{I,J=1}^{N}\sum_{r=1}^{\infty}\frac{1}{r}a^{I}_{-r,,L}D_{IJ}a^{J}_{-r,,R}}\nonumber\\&&\sum_{v^{I}}\cos\lbrack v^{I}(\varphi^{I}_{0,L}+(D\varphi_{0,R})^{I}+\alpha^{I})\rbrack|0\rangle. \end{eqnarray}
	Here the most relevant operators $\sum_{j=1}^{N}\cos\lbrack (\varphi^{I}_{0,L}+(D\varphi_{0,R})^{I}+\alpha^{I})/\sqrt{2}\rbrack$ in the cosine term are gapping potential which can condense the value of boson fields.
	If this interaction satisfies the Haldane condition~\cite{PhysRevLett.74.2090,wang2013non,wang2022symmetric,PhysRevLett.128.185301} for Tomonaga-Luttinger liquids:
	\begin{eqnarray}
	e^{T}_{I}K e_{J}=0,
	\label{con:Haldane}
	\end{eqnarray}
	then it can gap the compact boson CFT when the coupling constant is large. 
	
	In the open string picture, the Cardy state corresponds to the following boundary condition:
	\begin{eqnarray}
	\varphi^{I}_{L}=(D\varphi_{R})^{I}+\alpha^{I}.
	\end{eqnarray}
	So the computation above just implies the information on gapping potential from Cardy states or boundary conditions.
	
	Now for our interest, the boundary states on $t=0$ and $t=L$ are assumed to be $|\lbrace\alpha_{1}^{I}\rbrace,D_{1}\rangle$,  $|\lbrace\alpha_{2}^{I}\rbrace,D_{2}\rangle$. These two boundary states  are related by a symmetry transformation $g\in G$:
	\begin{eqnarray}
	|\lbrace\alpha_{2}^{I}\rbrace,D_{2}\rangle=g |\lbrace\alpha_{1}^{I}\rbrace,D_{1}\rangle.
	\end{eqnarray}
	
	Firstly we assume these two boundary stats have the same $D$ matrix.
	Then the amplitude is given by:
	\begin{eqnarray}
	Z(q)=(\eta(q))^{-N}\sum q^{u^{2}}\exp(-\frac{\beta}{8\pi L}(\alpha_{2}-\alpha_{1})^{2}),
	\end{eqnarray}
	where $u$ is in the Bravais lattice and $\eta(q)$ is the Dedekind function. Only the $\alpha_{2}=\alpha_{1}$, the partition survives in the limit $L$ $\to$ 0.  
	
	For the boundary states with the different $D$ matrices,  we can assume the number of eigenvalues $-$1 of $D_{1}D_{2}$ is $k$ ($k>0$) in the Ishibashi condition. The partition function is 
	\begin{eqnarray}
	Z(q)\propto (\eta(q))^{-N}(\mathscr{\theta}_{2}(q))^{k}\sum q^{u^{2}}.
	\end{eqnarray}
	It is easy to check that the partition function goes to zero in the limit $q \to$ 0 if the two boundary states have different $D$ matrices.
	
	Therefore we conclude only $\alpha_{2}=\alpha_{1}$ and $D_{2}=D_{1}$, which is equivalent to the boundary state $|A\rangle$ is symmetric, the partition function goes to 1 in the limit $L\to 0.$ This property of the partition function is just the special case of the anomaly free condition of our
	argument in the previous section. 
	 
	\subsection{Anomaly of U(1)  $\times$ $\mathbb{Z}_{2}$ symmetry}
	A simple example is the compact boson CFT with U(1) and $\mathbb{Z}_{2}$ symmetry. The  element in U(1) and  $\mathbb{Z}_{2}$ symmetry acts on
	the $\phi$ with radius $R=\sqrt{2}$:
	\begin{eqnarray}
	&& h^{-1}\phi h=\phi+\pi R, \nonumber\\
	&& h^{-1}\theta h=\theta+\pi R,\nonumber\\
	&&U^{-1} \theta U=\theta+\delta\theta,\nonumber\\
	&&U^{-1} \phi U=\phi\ .
	\end{eqnarray}
	
	On the lattice, this boson theory can be realized as the antiferromagnetic Heisenberg spin-1/2 chain. The U(1) symmetry is spin rotation
	on the $z$-axis and $\mathbb{Z}_{2}$ is the IR emergent $\mathbb{Z}_2$ partner of translation symmetry~\cite{Affleck:1988zj,OYA1997}.
	The LSM theorem claims that there is no gapped ground state keeping spin rotation on $z$-axis and translation symmetry if the spin in the unit cell is half-integer.
	As claimed above, if one can find a symmetric gapping potential with a unique ground state, there will exist symmetry invariant boundary states. Therefore we expect there will be symmetric boundary states only if we consider even copies of compact boson CFTs.
	
	For one copy of boson theory, the boundary state is just the Dirichlet boundary state or Neumann boundary state:
	\begin{eqnarray}
	|B,\phi_{0}\rangle_{D}=\sum_{v\in\frac{\sqrt{2}}{2}\mathbb{Z}} e^{iv\phi_{0}}|v\rangle\rangle_{D},\nonumber\\
	|B,\theta_{0}\rangle_{N}=\sum_{\nu\in\frac{\sqrt{2}}{2}\mathbb{Z}} e^{iv\theta_{0}}|v\rangle\rangle_{N}.
	\end{eqnarray}
	Here
	\begin{eqnarray}
	&&|v\rangle\rangle_{D}=e^{\sum_{r>0}\frac{1}{r}a_{-r,L}a_{-r,R}}|a_{0,L}=a_{0,R}=v\rangle,\nonumber\\
	&&|v\rangle\rangle_{N}=e^{\sum_{r>0}-\frac{1}{r}a_{-r,L}a_{-r,R}}|a_{0,L}=-a_{0,R}=v\rangle\nonumber\\
	\end{eqnarray} 
	are Ishibashi states satisfying the conformal boundary condition.
	
	Each boundary state can not be symmetric  since the  U(1) $\times$ $\mathbb{Z}_{2}$ symmetry changes both $\phi_{0}$ and $\theta_{0}$. 
	
	However, for two copies of the above theory, U(1) $\times$ $\mathbb{Z}_{2}$ symmetry is expected to be anomaly-free. The variables of two copies theory are noted by ($\phi_{1}$,$\theta_{1}$), ($\phi_{1}$,$\theta_{2}$). 

	We can construct a symmetric boundary state satisfying the constraint: 
	\begin{eqnarray}
	(\theta_{1}-\theta_{2})|B\rangle=\alpha_{1}|B\rangle,\quad
	(\phi_{1}+\phi_{2})|B\rangle=\alpha_{2}|B\rangle.
	\end{eqnarray}
	
	For simplicity, we redefine the boson fields \cite{han2017boundary}:
	\begin{eqnarray}
	\Phi_{1}=\frac{1}{\sqrt{2}}(\theta_{1}-\theta_{2}),\quad
	\Phi_{2}=\frac{1}{\sqrt{2}}(\phi_{1}+\phi_{2}).
	\end{eqnarray}
	The mode expansion is given by:
	\begin{eqnarray}
	\Phi_{i,L}=\Phi_{i,0,L}+\frac{2\pi}{L}(t+x)c_{i,0}+\sum_{r\ne 0} c_{i,r}e^{-\frac{2\pi i}{L}(t+x)},\nonumber\\
	\Phi_{i,R}=\Phi_{i,0,R}+\frac{2\pi}{L}(t-x)\bar{c}_{i,0}+\sum_{r\ne 0} \bar{c}_{i,r}e^{-\frac{2\pi i}{L}(t-x)}.
	\end{eqnarray}
	The symmetric boundary state for the redefined boson variables is:
	\begin{eqnarray}
	|B\rangle=(\frac{1}{\mathscr{N}_{N}}\sum_{n_{1}\in\mathbb{Z}}e^{iv_{1}\alpha_{1}} |v_{1}\rangle\rangle_{N})\otimes (\frac{1}{\mathscr{N}_{D}}\sum_{v_{2}\in\mathbb{Z}}e^{in_{2}\alpha_{2}} |v_{2}\rangle\rangle_{D}).\nonumber\\
	~
	\end{eqnarray}
	And the symmetric gapping potential is
	\begin{eqnarray}
	H'=U\lbrack\cos(\frac{1}{R}(\theta_{1}-\theta_{2}+\alpha_{1}))+\cos(\frac{1}{R}(\phi_{1}+\phi_{2}+\alpha_{2}))\rbrack. \nonumber\\
	~
	\end{eqnarray}For such gapping potential, it satisfies the Haldane condition (\ref{con:Haldane}) under which $H'$ is guaranteed to gap the bosons if $U$ is large.

	\subsection{Anomaly of $\mathrm{SU}(2)$  $\times$ $\mathbb{Z}_{2}$ symmetry}
	When the free boson CFT is on the self-dual point, i.e., the radius is $\sqrt{2}$, the U$(1)\times \mathbb{Z}_{2}$ symmetry can be extended into the SU(2)$\times$ $\mathbb{Z}_{2}$ symmetry and it is equivalent to $\mathrm{SU}(2)_{1}$ WZW model. Here the $\mathbb{Z}_{2}$ is still the symmetry  mentioned before and SU(2) symmetry is vector rotation subgroup of $\mathrm{SU}(2)_{L}$$\times$ $\mathrm{SU}(2)_{R}$ with generators density
	\begin{equation}
 \begin{split}
	\hat{S}^{z}&=\partial_{x}\phi,\\
	\hat{S}^{+}&=\exp(i\frac{1}{\sqrt{2}}(\phi+\theta))+\exp(i\frac{1}{\sqrt{2}}(-\phi+\theta)),\\
	\hat{S}^{-}&=\exp(-i\frac{1}{\sqrt{2}}(\phi+\theta))+\exp(-i\frac{1}{\sqrt{2}}(-\phi+\theta)).
 \end{split}
	\end{equation}
	On the lattice, SU(2) symmetry is the projective representation of SO(3) rotation and $\mathbb{Z}_{2}$ symmetry is still translation symmetry.
	
	From the previous analysis, there does not exist boundary states invariant under $\mathrm{SU}(2)\times$$\mathbb{Z}_{2}$ symmetry in the compact boson CFT.
	
	Similarly, for two copies of the compact boson CFT, there is no anomaly for SU(2)$\times$$\mathbb{Z}_{2}$ symmetry. Thus we expect there will be a symmetric boundary state. Now the generators of  SU(2) vector rotation are 
	\begin{eqnarray}
	\hat{S}^{z}&&=\partial_{x}(\phi_{1}+\phi_{2}),\nonumber\\
	\hat{S}^{+}&&=\sum_{i=1}^{2}\exp(i\frac{1}{\sqrt{2}}(\phi_{i}+\theta_{i}))+\sum_{i}\exp(i\frac{1}{\sqrt{2}}(-\phi_{i}+\theta_{i})),\nonumber\\
	\hat{S}^{-}&&=\sum_{i=1}^{2}\exp(-i\frac{1}{\sqrt{2}}(\phi_{i}+\theta_{i}))+\sum_{i}\exp(i\frac{1}{\sqrt{2}}(\phi_{i}-\theta_{i})).\nonumber\\
	\end{eqnarray}
	
	The symmetric Cardy states satisfy the constraint:
	\begin{eqnarray}
	&&\hat{S}^{z}|B\rangle=0,\quad
	\hat{S}^{+}|B\rangle=0,\nonumber\\
	&&\hat{S}^{-}|B\rangle=0,\quad
	h|B\rangle=|B\rangle.
	\end{eqnarray}
	A special solution of these equations is
	\begin{eqnarray}\label{bstate1}
	|B\rangle=(\frac{1}{\mathscr{N}_{N}}\sum_{v_{1}\in\frac{\sqrt{2}}{2}\mathbb{Z}} |v_{1}\rangle\rangle_{N})\otimes (\frac{1}{\mathscr{N}_{D}}\sum_{v_{2}\in\frac{\sqrt{2}}{2}\mathbb{Z}}e^{iv_{2}\sqrt{2}\pi} |v_{2}\rangle\rangle_{D}).\nonumber\\
	~
	\end{eqnarray}
	Here $|v_{1}\rangle\rangle_{N}$ and $|v_{2}\rangle\rangle_{D}$ are  the same  Ishibashi state mentioned before.
	The corresponding gapping potential is 
	\begin{eqnarray}
	H'=U(\cos(\frac{\sqrt{2}}{2}(\theta_{1}-\theta_{2}))+\cos(\frac{\sqrt{2}}{2}(\phi_{1}+\phi_{2}+\sqrt{2}\pi))\label{con:inter}.\nonumber\\
	\end{eqnarray}
	Moreover, it has been shown that the ground state is a trivial state when $U$ is positive whereas it is the Haldane state when $U$ is negative\cite{PhysRevB.93.104425,PhysRevLett.114.177204,PhysRevB.91.155150}.
	
	On the lattice, the two copies of the compact boson CFT on self-dual point can be realized as the antiferromagnetic SU(2) spin-1/2 ladder in 1+1 dimension:
	\begin{eqnarray}
	H=J\sum_{i} \vec{S}^{1}_{i}\cdot \vec{S}^{1}_{i+1}+\vec{S}^{2}_{i}\cdot \vec{S}^{2}_{i+1},\quad J>0.
	\end{eqnarray}
	Here $\vec{S}$ is the spin-1/2 operator. In the continuum limit, the spin operators can be related with effective low energy field operators as follows\cite{PhysRevD.11.3026,PhysRevD.11.2088,Affleck:1988zj,LUKYANOV1997571,10.1093/acprof:oso/9780198525004.001.0001,senechal2004introduction}:
	\begin{equation}
 \begin{split}    
	&\vec{S}(x)\approx \vec{J}(x)+(-1)^{x/a}\frac{\lambda}{\pi a}\vec{n}(x),\\
	&\vec{n}(x)=\left(\cos(\frac{\sqrt{2}}{2}\theta),\sin(\frac{\sqrt{2}}{2}\theta),\sin(\frac{\sqrt{2}}{2}\phi)\right),
 \end{split}
	\end{equation}
	where $\vec{J}$ and $\vec{n}$ are smooth and staggered parts and $x=ia$ and $a$ is lattice spacing.
 
	The gapping potential  (\ref{con:inter}) can be realized as the following interaction on the lattice:
	\begin{frame}
		
		\begin{tikzpicture}
		\node[label={[black]above:{$S^1_i$}}](b1) at (0, 0) [circle, draw=black!150, fill=black!90, text=blue!70] {};
		\node[label={[black]above:{$S^1_{i+1}$}}](b2) at (2, 0) [circle, draw=black!150, fill=black!90, text=blue!70] {};
		\node[label={[black]}](b3) at (4, 0) [circle, draw=black!150, fill=black!90, text=blue!70] {};
		\node[label={[black]}](b4) at (6, 0) [circle, draw=black!150, fill=black!90, text=blue!70] {};
		\node[label={[black]below:{$S^2_i$}}](b5) at (0, -2) [circle, draw=black!150, fill=black!90, text=blue!70] {};
		\node[label={[black]below:{$S^2_{i+1}$}}](b6) at (2, -2) [circle, draw=black!150, fill=black!90, text=blue!70] {};
		\node[label={[black]}](b7) at (4, -2) [circle, draw=black!150, fill=black!90, text=blue!70] {};
		\node[label={[black]}](b8) at (6, -2) [circle, draw=black!150, fill=black!90, text=blue!70] {};
		\node (c1) at (-1,0){};
		\node (c2) at (7,0){};
		\node (c3) at (-1,-2){};
		\node (c4) at (7,-2){};
		\node (c5) at (0,-0.4){};
		\node (c6) at (2,-0.4){};
		\node(c7) at(1.2,0){};
		\node(c8) at(0.8,0){};
		\node (c9) at (0,-1.6){};
		\node (c10) at (2,-1.6){};
		\node(c11) at(1.2,-2){};
		\node(c12) at(0.8,-2){};
		\node (c13) at (4,-0.4){};
		\node(c14) at(3.2,0){};
		\node(c15) at(2.8,0){};
		\node (c16) at (4,-1.6){};
		\node(c17) at(3.2,-2){};
		\node(c18) at(2.8,-2){};
		\node (c19) at (6,-0.4){};
		\node(c20) at(5.2,0){};
		\node(c21) at(4.8,0){};
		\node (c22) at (6,-1.6){};
		\node(c23) at(5.2,-2){};
		\node(c24) at(4.8,-2){};
		\node at(1,0.4) {$J$};
		\node at(1,-2.4) {$J$};
		\node at(-0.3,-1) {$J_{\bot}$};
		\node at(2.3,-1) {$J_{\bot}$};
		\node at(1,-0.3) {$J_{X}$};
		\node at(1,-1.7) {$J_{X}$};
		\path[-,line width=2pt] (b1) edge (b2);
		\path[-,line width=2pt] (b1) edge (c1);
		\path[-,line width=2pt] (b2) edge (b3);
		\path[-,line width=2pt] (b3) edge (b4);
		\path[-,line width=2pt] (b4) edge (c2);
		\path[-,line width=2pt] (b5) edge (b6);
		\path[-,line width=2pt] (b6) edge (b7);
		\path[-,line width=2pt] (b7) edge (b8);
		\path[-,line width=2pt] (b5) edge (c3);
		\path[-,line width=2pt] (b8) edge (c4);
		\path[-,line width=2pt] (b5) edge (b1);
		\path[-,line width=2pt] (b6) edge (b2);
		\path[-,line width=2pt] (b7) edge (b3);
		\path[-,line width=2pt] (b8) edge (b4);
		\draw[-,bend left,red] (c5) to [out=45,in=180] (c7);
		\draw[-,bend left,red] (c6) to [out=-45,in=180] (c8);
		\draw[-,bend left,red] (c9) to [out=-45,in=180] (c11);
		\draw[-,bend left,red] (c10) to [out=45,in=180] (c12);
		\draw[-,bend left,red] (c6) to [out=45,in=180] (c14);
		\draw[-,bend left,red] (c13) to [out=-45,in=180] (c15);
		\draw[-,bend left,red] (c10) to [out=-45,in=180] (c17);
		\draw[-,bend left,red] (c16) to [out=45,in=180] (c18);
		\draw[-,bend left,red] (c13) to [out=45,in=180] (c20);
		\draw[-,bend left,red] (c19) to [out=-45,in=180] (c21);
		\draw[-,bend left,red] (c16) to [out=-45,in=180] (c23);
		\draw[-,bend left,red] (c22) to [out=45,in=180] (c24);
		
		\end{tikzpicture}
		
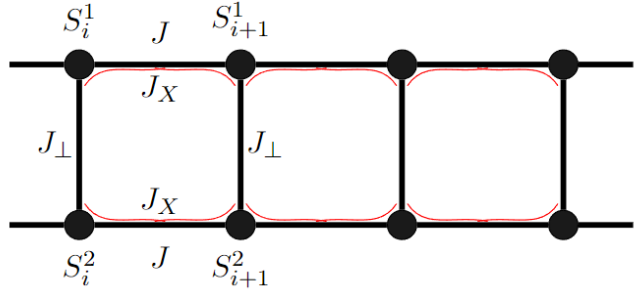
\captionof{figure}{\textbf{SU(2) and translation invariant gapping potential.}}\label{Fig.gap potential}
	\end{frame}

	\begin{eqnarray}
	H'=U\lambda_{1}^{2}\sum_{i}\vec{S}^{1}_{i}\cdot \vec{S}^{2}_{i}-U\lambda_{2}^{2}\sum_{i}(\vec{S}^{1}_{i}\cdot \vec{S}^{1}_{i+1})(\vec{S}^{2}_{i}\cdot \vec{S}^{2}_{i+1}),\nonumber\\
	\end{eqnarray}
	, which is shown in Fig. \ref{Fig.gap potential}.
 Here    $\lambda_{1}$ and $\lambda_{2}$ are nonuniversal constants\cite{affleck1998exact}. On the lattice, the SU(2) $\times$ $\mathbb{Z}_{2}$ symmetry is spin rotation and IR emergent $\mathbb{Z}_2$ partner of translation symmetry. This interaction is invariant under the spin rotation and translation transformation. To see this lattice model having a unique gapped ground state, we can do the Jordan Wigner transformation and this potential will give a unique ground state and fermionic mass $m\propto 3U$  \cite{chen2017model}. This result agrees with the LSM theorem which claims there will be a gapped ground state keeping SO(3) and translation symmetry if the spin in each unit cell is integer.
	
Indeed, the standard Abelian bosonization approach~\cite{PhysRevB.34.6372} to Haldane gap problem is based on
	the description of the spin-$S$ antiferromagnetic chain as $2S$ coupled spin-$1/2$ chains, which is mapped to
	$2S$-component boson field theory (Tomonaga-Luttinger Liquid) with possible interactions.
	Often we can focus on one particular linear combination corresponding to the ``center-of-mass'' field, in the low energy limit.
	However, when $2S$ is even,
	the translation symmetry acts trivially on this center-of-mass field and thus we do not expect anomalies.
	Thus we generically expect the ``Haldane gap''.
	In Ref.~\onlinecite{cho2017relationship}, the degeneracy in the BCFT spectrum similar to what we discuss in this paper
	was used as a probe of the Haldane SPT phase. In their construction, the boundary conditions imposed on the two sides
	correspond to different phases (trivial and SPT), and are not related by a symmetry transformation of the system.
	Despite these differences, it would be interesting to find deeper connections between their analysis and ours.

	\subsection{Anomaly of $\mathrm{SU}(N)$  $\times$ $\mathbb{Z}_{N}$ symmetry}
	The discussion on the $\mathrm{SU}(2)_{1}$ WZW model can be generalized to $\mathrm{SU}(N)_{1}$ WZW model with $\mathrm{SU}(N)$ $\times$ $\mathbb{Z}_{N}$ symmetry. For $\mathrm{SU}(N)_{1}$WZW model, the generators of vector $\mathrm{SU}(N)$ transformation is:
	\begin{eqnarray}
	H^{i}&=&i\partial_{x}\phi^{i},\nonumber\\
	E^{\alpha}&=&e^{i\alpha\cdot \varphi_{L}}+e^{-i\alpha\cdot \varphi_{R}},
	\end{eqnarray}
	where $\alpha$ are root vectors with $N-1$ components, $H$ are generators of the Cartan subgroups and $E$ are other vertex generators. Since the $\mathrm{SU}(N)$ symmetry has $N-$1 $H$ generators, the $\mathrm{SU}(N)_{1}$ WZW model can be described by $N-1$ free massless bosons. We should note here the radius of bosons is not $\sqrt{2}$ but depends on the root of  $\mathrm{SU}(N)$ Lie algebra.
	
	The $\mathbb{Z}_{N}$ symmetry only acts on the last massless boson:
	\begin{eqnarray}
	&&g \varphi^{N-1}_{L} g^{-1}=\varphi^{N-1}_{L}+4\pi\sqrt{\frac{N-1}{N}},\nonumber\\
	&& g \varphi^{N-1}_{R} g^{-1}=\varphi^{N-1}_{R}.
	\end{eqnarray}
	
	The condition for symmetric boundary states is:
	\begin{eqnarray}
	&&H^{i}|B\rangle=0,\quad
	E^{\alpha}|B\rangle=0,\nonumber\\
	&&g|B\rangle=|B\rangle.
	\end{eqnarray}
	
	For one copy of the $\mathrm{SU}(N)_{1}$ WZW model,  in analogy to the previous analysis for the SU(2) case, there does not exist a symmetric boundary state.
	
	However, as the index of LSM anomaly shows, one can find a trivial gapped ground state for $N$ copies of  $\mathrm{SU}(N)_{1}$ WZW model \cite{Yao-Hsieh-Oshikawa19}. More precisely, there exists $\mathrm{SU}(N)$ and translational invariant gapped interaction with the unique ground state. Therefore we expect to construct a symmetric boundary state for $N$ copies of  $\mathrm{SU}(N)_{1}$ WZW model.

	Firstly, it is natural to ask how many there are  independent constraints for boundary states. To solve this question, we first notice when $\alpha$ is a root, then $-\alpha$ is also a root. Therefore, we can only consider positive roots where its first non-zero component is positive. Another helpful information is the communication relations between vertex generators with positive roots:
	\begin{eqnarray}
	&&[E^{\alpha},E^{\beta}]\propto \frac{E^{\alpha+\beta}}{x_{1}-x_{2}}\ , \quad \alpha+\beta\in\Delta_{+},\nonumber \\
	&&\lbrack E^{\alpha},E^{\beta}\rbrack=0\ ,\quad \alpha+\beta\notin\Delta_{+},
	\end{eqnarray}
	where $\Delta_{+}$ is the set of positive roots. As a result, we can focus on independent positive roots that can not be written as the sum or minus of two different positive roots. Actually, there are $N-$1 independent positive roots for $\mathrm{SU}(N)$ group:
	\begin{eqnarray}\label{bstate2}
	&& m_{1}=\frac{\sqrt{2}}{2}(1,0,0,\cdots,0)\ ,\nonumber\\
	&&m_{2}=\frac{\sqrt{2}}{2}(\frac{1}{2},\frac{\sqrt{3}}{2},0,\cdots,0),\nonumber\\
	&&\vdots\nonumber\\
	&~&m_{k}=\frac{\sqrt{2}}{2}(\frac{1}{2},\frac{1}{2\sqrt{3}},\cdots,\sqrt{\frac{1}{2k(k-1)}},\sqrt{\frac{k+1}{2k}},0,\cdots,0),\nonumber\\
	&&\vdots\nonumber\\
	&~&m_{N-1}=\frac{\sqrt{2}}{2}\left(\frac{1}{2},\frac{1}{2\sqrt{3}},\cdots,\sqrt{\frac{N}{2(N-1)}}\right).
	\end{eqnarray}
	
	So for $N$ copies of  $\mathrm{SU}(N)_{1}$ WZW model, the independent constraints for symmetric boundary states can be written in the form of vertex operators with $m$ vector:
	\begin{eqnarray}
	&&\sum_{j=1}^{N} E^{m}_{j}|B\rangle=0,\quad
	\sum_{j}^{N}H^{i}_{j}|B\rangle=0,\nonumber\\
	&&h|B\rangle=|B\rangle.
	\end{eqnarray}
	Here $j$ is the index for the $j$th copy.
	
	We find a special solution for these equations:
	\begin{eqnarray}
	&&\varphi^{1}_{j,L}+\varphi^{1}_{j+1,R}=\sqrt{2}\pi,\nonumber\\
	&&\varphi^{2}_{j,L}+\varphi^{2}_{j+1,R}=\frac{2\pi}{\sqrt{6}},\nonumber\\
	&&\vdots\nonumber\\
	&&\varphi^{k}_{j,L}+\varphi^{k}_{j+1,R}=2\pi\sqrt{\frac{1}{(k+1)k}},\nonumber\\
	&&\vdots\nonumber\\
	&&\varphi^{N-1}_{j,L}+\varphi^{N-1}_{j+1,R}=2\pi\sqrt{\frac{1}{(N-1)N}}.  \end{eqnarray}

	\section{Anomaly of WZW model}\label{sec 5}
	In this section, we will discuss the WZW models with center symmetry, vector rotation symmetry and time reversal symmetry. We will show our argument is still valid for these models.

	\subsection{Introduction to WZW model and global symmetry}
	The WZW model with the  level $k$ is defined with the action:
	\begin{eqnarray}\label{wzw}
	S=&&\frac{k}{16\pi}\int_{M_{2}} dxdt\ Tr(\partial_{\mu}g^{-1}\partial^{\mu}g)\nonumber\\&&+\frac{k}{24\pi}\int_{B_{3}} d^{3}x\ Tr\lbrack(\tilde{g}^{-1}d\tilde{g})^{3}\rbrack,
	\end{eqnarray}
	where $g$ is a smooth map from the spacetime $M_2$ to the group manifold $G$ and the second term is defined on a 3-dimensional manifold $B_{3}$ whose boundary is the original spacetime $M_2$. $\tilde{g}$ is denoted to be an extension of $g$ field to $B_{3}$ and $k$ is a non-negative integer.
	
	There are three kinds of global symmetry: \\
	1. Center symmetry $\Gamma$ is the center of the group $G$ which acts as multiplying a U(1) phase:
	\begin{eqnarray}
	g\to hg,\quad h\in \Gamma.
	\end{eqnarray}
	2. Vector rotation symmetry acts as the adjoint representation of $G$:
	\begin{eqnarray}
	g\to VgV^{-1}\ ,\quad V\in G.
	\end{eqnarray}
	3. Charge conjugation symmetry $\mathcal{C}$ maps the group element to its complex conjugation (without transpose):
	\begin{eqnarray}
	\mathcal{C}:~ g\to g^*.
	\end{eqnarray}
	
	\subsection{Action of global symmetry on Cardy state}
	For the WZW model, there is a set of special solutions for the Cardy condition \cite{cardy2004boundary}:
	\begin{eqnarray}
	n^{a}_{\alpha \beta}=\sum_{i}\frac{S_{\alpha i}S_{\beta i}S_{ia}}{S_{0i}}\in \mathbb{Z}^{+},\quad
	|B_{\alpha}\rangle=\sum_{i}\frac{S_{\alpha i}}{\sqrt{S_{0i}}}|i\rangle\rangle,\nonumber\\
	\end{eqnarray}
	where $S_{\alpha i}$ is elements of $S$ matrix of the WZW model and $|i\rangle\rangle$ is Ishibashi state.
	
	The action of center symmetry on the Cardy states is the same as the action of outer automorphism $A$ on the Dynkin label. More precisely, the action of center symmetry on the Ishibashi state is given by
	\begin{eqnarray}
	h |i\rangle\rangle=e^{-2\pi(A\omega_{0}, i)}|i\rangle\rangle.
	\end{eqnarray}
	Then we can obtain its action on Cardy states\cite{numasawa2018mixed}:
	\begin{eqnarray}
	h |B_{\alpha}\rangle&&=\sum_{i}\frac{S_{\alpha i}}{\sqrt{S_{0i}}}e^{-2\pi(A\omega_{0}, i)}|i\rangle\rangle\nonumber\\&&=\sum_{i}\frac{(AS)_{\alpha i}}{\sqrt{S_{0i}}}|i\rangle\rangle=|B_{A\alpha}\rangle.
	\end{eqnarray}
	Here we use the relationship:
	\begin{eqnarray}
	(AS)_{\alpha i}=S_{\alpha i}e^{-2\pi(A\omega_{0}, i)}.
	\end{eqnarray}
	
	The charge conjugation $\mathcal{C}$ will act on the Cardy states as follows:
	\begin{eqnarray}
	\mathcal{C} |B_{\alpha}\rangle&=&\sum_{i}\frac{S_{\alpha i}}{\sqrt{S_{0i}}}|i^{+}\rangle\rangle,\nonumber\\&=&C_{\alpha\beta}|B_{\beta}\rangle,
	\end{eqnarray}
	where $C$ matrix is the charge conjugation transformation of the Dynkin label.

	\subsection{Mixed anomaly of WZW model}

	\subsubsection{Invarinat boundary states  of center symmetry and vector rotation symmetry}
	Let's firstly consider $\mathrm{SU}(N)_{k}$ WZW model with the center symmetry $ \mathbb{Z}_{N}$ and vector $\mathrm{PSU}(N)$ symmetry.

	The generators of $\mathrm{PSU}(N)$ symmetry can be written in the chiral form:
	\begin{eqnarray}
	S^{a}=J^{a}_{0}+\bar{J}^{a}_{0}.
	\end{eqnarray}
	Here $J^{a}_{0}$ and $\bar{J}^{a}_{0}$ are zero modes of holomorphic and antiholomorphic affine Lie group.
	
	As we discussed in the previous section, if the WZW model is anomaly free with respect to  $\mathrm{PSU}(N)$ $\times$ $\mathbb{Z}_{N}$ symmetry, we expect to construct the symmetric boundary state:
	\begin{eqnarray}
	V|B\rangle=|B\rangle,\quad
	h|B\rangle=|B\rangle.
	\end{eqnarray}
	
	To construct such a boundary state, we should know how the $\mathrm{PSU}(N)$ symmetry acts on the Ishibashi state. In analogy to the boundary state of compact boson CFTs, we denote an orthogonal basis in the holomorphic and antiholomorphic sectors:
	\begin{eqnarray}\label{eq:basis}
	|\phi^{\lambda}_{i},m\rangle=\prod_{k=1,a}\frac{1}{\sqrt{m_{k}}!}(\frac{j^{a}_{-k}}{\sqrt{k}})^{m_{k}}|\phi^{\lambda}_{i},0\rangle,\nonumber\\
	|\bar{\phi}^{\lambda}_{i},m\rangle=\prod_{k=1,a}\frac{1}{\sqrt{m_{k}}!}(\frac{\bar{j}^{a}_{-k}}{\sqrt{k}})^{m_{k}}|\bar{\phi}^{\lambda}_{i},0\rangle.
	\end{eqnarray}
	Here $|\phi^{\lambda}_{i},0\rangle$ is vacua in different positive representations $\lambda$ of the Virasoro algebra. Now we can construct the Ishibashi state:
	\begin{eqnarray}\label{Ishibashi}
	|\lambda\rangle\rangle=\sum_{m}|\phi_{i}^{\lambda},m\rangle\otimes U|\bar{\phi}^{\lambda}_{i},\bar{m}\rangle,
	\end{eqnarray}
	where $U$ operator is an anti-unitary operator acting on the chiral generators as follows:
	\begin{eqnarray}
	U^{-1} \bar{J}^{a}_{n} U=-(\bar{J}^{a}_{-n})^{+}=-(\bar{J}^{a}_{n}).
	\end{eqnarray}
	
	The basis \eqref{eq:basis} forms an irreducible representation of vector $\mathrm{PSU}(N)$ symmetry:
	\begin{eqnarray}
	J^{a}|\phi^{\lambda}_{i},m\rangle=\sum_{j} T^{\lambda,a}_{ij}|\phi^{\lambda}_{j},m\rangle,\nonumber\\
	\bar{J}^{a}|\bar{\phi}^{\lambda}_{i},m\rangle=\sum_{j} T^{\lambda,a}_{ij}|\bar{\phi}^{\lambda}_{j},m\rangle.
	\end{eqnarray}
	Here $T^{\lambda,a}_{ij}$ are generators of $\mathrm{SU}(N)$ in the $\lambda$ representation.
	
	From the above equation, we can show all the Ishibashi states are invariant under vector $\mathrm{PSU}(N)$ transformation. The detail is shown in Appendix~\ref{app_B}.
	As a result, the Cardy state is also invariant under vector $\mathrm{PSU}(N)$ transformation, which means we only need to find Cardy states satisfying the following conditions
	\begin{eqnarray}
	h|B\rangle=|B\rangle.
	\end{eqnarray}
	
	If we denote the Cardy state using  the  affine  Dynkin  labels  [$\lambda_{0}$;$\lambda_{1}$,$\cdots$,$\lambda_{N-1}$], the action of the center symmetry is given by
	\begin{eqnarray}
	A:[\lambda_{0};\lambda_{1},\cdots,\lambda_{N-1}]\to [\lambda_{N-1};\lambda_{0},\cdots,\lambda_{N-2}].
	\end{eqnarray}
	The solution for invariant condition is:
	\begin{eqnarray}
	&&\lambda_{0}=\lambda_{1}=\cdots=\lambda_{N-1},\\
	&&k=\lambda_{0}+\lambda_{1}+\cdots+\lambda_{N-1}=N\lambda_{0}.
	\end{eqnarray}  
	Therefore only when $k$ is a multiple of $N$, there exists a symmetric Cardy state. Since the index of mixed anomaly with respect to $\mathrm{PSU}(N)$ $\times$ $\mathbb{Z}_{N}$ symmetry is $k$ mod $N$,  there exists a translation and $\mathrm{PSU}(N)$ symmetric gapped ground state only when the level $k$ is a multiple of $N$ \cite{Yao-Hsieh-Oshikawa19}. So the result of the BCFT approach agrees with this classification of mixed anomaly and the symmetric Cardy state corresponds to the translation and $\mathrm{SU}(N)$ invariant ground state.
	
	Moreover, for the $\mathrm{SU}(2)_{2k}$ WZW model, we can show the symmetric boundary state corresponds to the massive phase after adding symmetric gapping potential $\lambda \tr(g^{2})$ ($\lambda>0$). The ground state is in the trivial phase when $k$ is even, while it is in the Haldane phase when $k$ is odd\cite{lecheminant2015massless}. This can also be seen from the partition function of an open string. The detail is shown in the Appendix~\ref{app d}.
	
	We can also apply this argument to the WZW model with other simple Lie algebras.
	The results are listed in TABLE~\ref{table1}. The detail is shown in the Appendix~\ref{app_B}. 
	The anomaly-free conditions on the level $k$ for most Lie algebras there implies the classification of LSM-type anomaly, e.g., $k\in N\mathbb{N}$ for SU$(N)$ consistent with the corresponding $\mathbb{Z}_N$ classification of LSM anomaly index~\cite{Yao-Hsieh-Oshikawa19} and the projective representation argument~\cite{PhysRevLett.126.217201} on lattices.

Moreover, our results for the Spin$(n)$ WZW models including $B$ and $D$ series are also consistent with a generalized LSM theorem for SO$(n)$ spin chains proposed in Ref.~\onlinecite{Tu2011EffectiveFT, Jian-Bi-Xu18}, which claims that an SO$(n)$ spin chain in the $n$-dimensional vector representation with even $n$ has either gapless excitations or degenerate gapped ground states with broken translational symmetry. In particular, the model studied in Ref.~\onlinecite{Tu2011EffectiveFT} is the SO$(n)$ bilinear-biquadratic spin chain whose low-energy physics can be described by the $\mathrm{Spin}(n)_{1}$ WZW model. According to our results in TABLE~\ref{table1} (or TABLE \ref{table4}), such a low-energy theory indeed has an LSM-type anomaly if $n$ is even.

	\begin{table*}[t]
		\centering
		\begin{tabular}{c c c c c c }
			\hline\hline
			Cartan matrix \quad\quad	&Group\quad \quad   &Action of $h$\  \quad&Invariant Boundary states \\
			\hline
			$A_{N-1}$&SU($N$)&$\lbrack\lambda_{N-1};\lambda_{0},...,\lambda_{N-2}\rbrack$&$k\in N\mathbb{N}$\\
			\\[-1em]
			$B_{N}$&Spin($2N+1$)&$\lbrack\lambda_{1};\lambda_{0},..,\lambda_{N-1},\lambda_{N}\rbrack$&$k\in$$\mathbb{N}$\\
			\\[-0.7em]
			$C_{N}$&USp($N$)&$\lbrack\lambda_{N};\lambda_{N-1},...,\lambda_{1},\lambda_{0}\rbrack$&$k\in$$2\mathbb{N}$ or $N\in$$2\mathbb{N}$\\
			\\[-0.9em]
			$D_{2N+1}$&Spin($4N+2$)&$\lbrack\lambda_{2N};\lambda_{2N+1},\cdots,\lambda_{1},\lambda_{0}\rbrack$&$k\in$4$\mathbb{N}$ \\
			\\[-0.7em]
			$D_{2N}$&Spin($4N$)&$\lbrack\lambda_{1};\lambda_{0},\lambda_{2},\cdots,\lambda_{2N},\lambda_{2N-1}\rbrack$&$k\in$2$\mathbb{N}$\\
			\\[-0.7em]
			$D_{2N}$&Spin($4N$)&$\lbrack\lambda_{2N};\lambda_{2N-1},\lambda_{2N-2},\cdots,\lambda_{0}\rbrack$&--\\
			\\[-0.7em]
			$E_{6}$&	$E_{6}$&$\lbrack\lambda_{1};\lambda_{5},\lambda_{4},\lambda_{3},\lambda_{6},\lambda_{0},\lambda_{2}\rbrack$&$k\in$$3\mathbb{N}$\\
			\\[-0.7em]
			$E_{7}$&$E_{7}$&$\lbrack\lambda_{6};\lambda_{5},\lambda_{4},\lambda_{3},\lambda_{2},\lambda_{1},\lambda_{0},\lambda_{7}\rbrack$&$k\in$$2\mathbb{N}$\\
			\\[-1em]
			
			\\[-1em]
			\hline\hline
		\end{tabular}
		\caption{The action of the center symmetry on Cardy state in the WZW model and  condition for the existing of symmetric Cardy state.  }\label{table1}
	\end{table*}
	\subsubsection{Invariant boundary states of time reversal and vector rotation symmetry}
Besides, we can also consider WZW models with vector rotation and time-reversal symmetries. There are two types of time-reversal symmetry. One acts as $\mathcal{T}_{1}$: $g(x,t) \to  g^{-1}(x,-t)$ and the other is given by the combination of $\mathcal{T}_{1}$ and any order 2 element $h_2$ (if exists) of the center symmetry group, $\mathcal{T}_{2}=\mathcal{T}_{1}h_2$: $g(x,t) \to -g^{-1}(x,-t)$; either one squares to the identity, i.e. $\mathcal{T}_{1}^2=\mathcal{T}_{2}^2=1$. As we will see in the following discussion, however, it is only $\mathcal{T}_{2}$ which has a mixed anomaly with the vector rotation symmetry.
	
To construct the symmetric Cardy states, we work in Euclidean
signature by performing the Wick rotation $t=-i\tau$. We should also consider the boundary condition in the closed channel where the space-time cylinder has been rotated by $\pi/2$, namely, $(x',\tau')$ = $(\tau,-x)$.  Due to the Lorentz symmetry of CFTs, the original time-reversal symmetry $\mathcal{T}_{\eta}$, which is an anti-unitary operator in the
Lorentz signature, becomes the unitary $\mathcal{CP}_{\eta}$ symmetry in the Euclidean signature:
\begin{eqnarray}
&&\mathcal{CP}_{1}: g(x',\tau')\to (g^{*})^{-1}(-x',\tau'),\\
&&\mathcal{CP}_{2}: g(x',\tau')\to-(g^{*})^{-1}(-x',\tau').
	\end{eqnarray}
	Here we can see $\mathcal{CP}_{1}$ is the combination of charge conjugation $\mathcal{C}$ and  spatial reflection $\mathcal{P}$.
	
	In the first case, the Cardy state should satisfy the constraint as follows:
	\begin{eqnarray}\label{conjugation}
	\mathcal{CP}_{1}|B\rangle=|B\rangle,\quad
	V|B\rangle=|B\rangle.
	\end{eqnarray}
	Since the  spatial reflection maps $x'$ to $-x'$, it exchanges the state of holomorphic and antiholomorphic sections. Thus the 
	Ishibashi states (\ref{Ishibashi}) and Cardy states are invariant under the  spatial reflection symmetry.
	
	Thus we only need to find the invariant Cardy state under the charge conjugation. In other words, the Dynkin index of such Cardy states should be invariant under the following transformation:
	\begin{eqnarray}
	C:[\lambda_{0};\lambda_{1},\cdots,\lambda_{N-1}]\to [\lambda_{0};\lambda_{N-1},\cdots,\lambda_{2},\lambda_{1}].
	\end{eqnarray}
	The roots of boundary state should satisfy :
	\begin{eqnarray}
	\lambda_{i}=\lambda_{N-i}, \quad i>0.
	\end{eqnarray}
	When $N$ is even, $k$ is given by
	\begin{eqnarray}
	k=\lambda_{0}+2(\lambda_{1}+\cdots+\lambda_{\frac{N}{2}-1})+\lambda_{\frac{N}{2}}. 
	\end{eqnarray}  
	On the other hand, when $N$ is odd, $k$ is given by
	\begin{eqnarray}
	k=\lambda_{0}+2(\lambda_{1}+\cdots+\lambda_{\frac{N-1}{2}}).
	\end{eqnarray}    
	In each condition, $k$ can take an arbitrary natural number. Therefore, there is no mixed anomaly between the $\mathcal{T}_{1}$ and $\mathrm{PSU}(N)$ symmetry.
	
	In the second case, since the $\mathcal{T}_2$ symmetry needs  an order 2 element of center symmetry, it only exists when $N$ is even.  The condition for an invariant boundary state is:
	\begin{eqnarray}
	&&\mathcal{CP}_{2}|B\rangle=\mathcal{CP}_{1}h_2|B\rangle=|B\rangle,\nonumber\\
	&&V|B\rangle =|B\rangle.
	\end{eqnarray} 
	Since all Cardy states are invariant under the  spatial reflection symmetry, we only need to find the invariant Cardy state under  the  combination of charge conjugation and the order 2 element of the center symmetry group.
	
	The Dynkin roots of such boundary state should satisfy
	\begin{eqnarray}
	[\lambda_{0};\lambda_{1},...,\lambda_{N-1}]= [\lambda_{N/2};\lambda_{N/2-1},\cdots,\lambda_{N/2+2},\lambda_{N/2+1}].\nonumber\\
	~
	\end{eqnarray} 
	Thus the relations $\lambda_{i}=\lambda_{N/2-i}$ if $0\leq i\leq N/2$ and $\lambda_{i}=\lambda_{3N/2-i}$ if $N/2< i\leq N-1$) hold for symmetric boundary states. When $N\in 4\mathbb{Z}+2$, the level $k$ is given by
	\begin{eqnarray}
	k=2(\lambda_{0}+\cdots+\lambda_{\frac{N-2}{4}}+\lambda_{\frac{N+2}{2}}+\cdots+\lambda_{\frac{3N-2}{4}}).
	\end{eqnarray}
 And when $N\in 4\mathbb{N_{+}}$, $k$ is given by
 \begin{eqnarray}
	k=&&	2(\lambda_{0}+\cdots+\lambda_{\frac{N-4}{4}}+\lambda_{\frac{N+2}{2}}+\cdots+\lambda_{\frac{3N-4}{4}})\nonumber\\
	&&+\lambda_{\frac{N}{4}}+\lambda_{\frac{3N}{4}}.
	\end{eqnarray}
	
	So a symmetric boundary state exists only when $k$ is even or when $N$ is a multiple of 4. As a result, there should be no mixed anomaly between $\mathrm{PSU}(N)$ and $\mathcal{T}_{2}$ symmetry under the same condition.

\subsection{Application to a spin chain with the triple product interactions}

On the other hand, when $N\in4\mathbb{N}+2$, $\mathrm{PSU}(N)$ and $\mathcal{T}_{2}$ have a mixed anomaly. This anomaly implies an LSM-type ingappability under the  $\mathrm{PSU}(N) \times \mathcal{T}_{2}$ symmetry.
Let us illustrate the point with the simplest case with $N=2$.
A lattice model which reflect the ingappability of the $\mathrm{SU}(2)_{1}$ WZW theory is given
by the spin-1/2 chain with SO(3)-invariant three-spin (triple-product) interactions \cite{Schmoll:2019aa} 
\begin{eqnarray}\label{triple-product}
H=\sum_j (-1)^{j}\bold{S}_{j}\cdot (\bold{S}_{j+1}\times\bold{S}_{j+2}) .
\label{t2}
\end{eqnarray} 
Although this model is invariant under the $\mathrm{PSU}(2)$ symmetry, it is invariant only under two-site translation and
not one-site translation. Thus, the standard LSM theorem does not give any constraint on this model.
Nevertheless, in Ref.~\onlinecite{Schmoll:2019aa} the system was found to be gapless and described by the
$\mathrm{SU}(2)_{1}$ WZW theory.
To our knowledge, the mechanism behind this phenomenon has not been clarified.
Here we point out that the gapless nature of the model is not accidental but rather reflects the ingappability of the
$\mathrm{SU}(2)_{1}$ WZW theory due to the mixed anomaly of $\mathrm{PSU}(2)$ and $\mathcal{T}_{2}$ symmetries.

On the lattice, $\mathcal{T}_2$ corresponds to the combination of the time reversal $\mathcal{T}_1: \bold{S}_j \to - \bold{S}_j$
and the one-site translation.
Thus, the mixed anomaly implies an LSM-type ingappability of one-dimensional
$\mathrm{PSU}(2)$-symmetric spin systems which are invariant under the combined operation.
This is a field-theory derivation of a simple one-dimensional case of the
LSM-type ingappability due to magnetic space group symmetries~\cite{Watanabe-MSG,Yao:2023bnj}.

Indeed, we can observe that the model~\eqref{t2} has this symmetry, since the Hamiltonian is odd under 
both the time reversal $\mathcal{T}_1$ and the one-site translation.
Therefore, the model~\eqref{t2} can be gapped only if the ground states are at least doubly degenerate.
This was a background why it was gapless without any fine-tuning of parameters\cite{Schmoll:2019aa}.

In fact, 
this ingappability does not require the full $\mathrm{PSU}(2)$ symmetry and
can be protected by the smaller symmetry
\begin{eqnarray}
&&\mathcal{T}_2: T_2=TKU_{T}=TK\prod^{j=L}_{j=1}i\sigma^y_j,\\
&&U(1)_z: U_{\theta}=\exp(i\theta\sum^{L}_{j=1}  \bold{S}^z_{j}) ,\\
&&\mathbb{Z}^y_2: R^{\pi}_{y}=\prod^{L}_{j=1}i\sigma^y_{j} .
\end{eqnarray}
We present a simple proof in Appendix~\ref{LSM}, which is analogous to the original proof of the LSM theorem.

Moreover, we can obtain a similar argument of the WZW models with other simple Lie groups. On the one hand, the result for $\mathcal{T}_{2}$  and vector rotation symmetry is listed in the Table~\ref{table2}. The detail of calculation is  shown in the Appendix~\ref{app_c}. On the other hand, there is always no mixed anomaly between $\mathcal{T}_{1}$ and vector rotation symmetry.

As mentioned in the Introduction, anomaly-free conditions of the WZW model were discussed earlier in Ref.~\onlinecite{numasawa2018mixed}
using a similar formulation.
However, they discuss the anomaly-free condition for the center symmetry with the charge conjugation.
In contrast, we have studied the mixed anomaly of the Lie Group and center symmetries, which is more relevant
to the LSM-type ingappability of lattice models.

	\begin{table*}[t]
		\centering
		\begin{tabular}{c c c c c c }
			\hline\hline
			Cartan matrix \quad\quad	&Group\quad \quad   &action of $CP_{2}$\  \quad&Invariant Boundary states \\
			\hline
			$A_{N-1}$&SU($N$)&$\lbrack\lambda_{N/2};\lambda_{N/2-1},...,\lambda_{N/2+1}\rbrack$&$k\in$$2\mathbb{N}$ or $N\in$$4\mathbb{N}$\\
			\\[-1em]
			$B_{N}$&Spin($2N+1$)&$\lbrack\lambda_{1};\lambda_{0},..,\lambda_{N-1},\lambda_{N}\rbrack$&$k\in$$\mathbb{N}$\\
			\\[-0.7em]
			$C_{N}$&USp($N$)&$\lbrack\lambda_{N};\lambda_{N-1},...,\lambda_{1},\lambda_{0}\rbrack$&$k\in$$2\mathbb{N}$ or $N\in$$2\mathbb{N}$\\
			\\[-0.9em]
			$D_{2N+1}$&Spin($4N+2$)&$\lbrack\lambda_{1};\lambda_{0},\lambda_{2},...,\lambda_{2N},\lambda_{2N+1}\rbrack$&$k\in$$\mathbb{N}$ \\
			\\[-0.7em]
			$D_{2N}$&Spin($4N$)&$\lbrack\lambda_{1};\lambda_{0},\lambda_{2},...,\lambda_{2N},\lambda_{2N-1}\rbrack$&$k\in$2$\mathbb{N}$\\
			\\[-0.7em]
			$D_{2N}$&Spin($4N$)&$\lbrack\lambda_{2N};\lambda_{2N-1},\lambda_{2N-2},...,\lambda_{1},\lambda_{0}\rbrack$&--\\
			\\[-0.7em]
			
			$E_{7}$&$E_{7}$&$\lbrack\lambda_{6};\lambda_{5},\lambda_{4},\lambda_{3},\lambda_{2},\lambda_{1},\lambda_{0},\lambda_{7}\rbrack$&$k\in$$2\mathbb{N}$\\
			\\[-1em]
			
			\\[-1em]
			\hline\hline
		\end{tabular}
		\caption{The action of $\mathcal{CP}_{2}$  symmetry on Cardy state in the WZW model and  condition for the existing of symmetric Cardy state.}\label{table2}
	\end{table*}
	
	\section{Discrete symmetries of minimal models}\label{sec 7}
	In this section, we will discuss the global discrete symmetries of minimal models.
	
 	The classification of $c<1$ minimal models is given in terms of a pair of Dynkin diagrams ($A_h$,$G$) where $G$ is $A-D-E$ type. The boundary states of ($A_{h}$,$G$) models are labeled by pair ($r$,$a$) of nodes of ($A_h$,$G$) graph with the identification\cite{Behrend1999BoundaryCI}:
 	\begin{eqnarray}
 	(r,a)=(h+1-r,\gamma(a)).
 	\end{eqnarray}
 	Here $\gamma$ is an automorphism of the $G$ graph. Thus independent boundary states are half of nodes of the ($A_h$,$G$) graph.
 	
 	These minimal models have a unique and maximal $\mathbb{Z}_2$ symmetry except six cases\cite{1998NuPhB.535..650R}. The ($A_{4}$,$D_4$) (the critical 3-state Potts model) and ($A_{6}$,$D_4$) (the tricritical 3-state Potts model) have a $\mathbb{Z}_2$ and $\mathbb{Z}_3$ symmetry which combine to an $S_3$ symmetry. The remaining four models related to the $E_7$ and $E_8$ have no symmetry.   
 	
 	\subsection{Invariant boundary states of $\mathbb{Z}_{2}$ symmetry}
 	Let us begin with the invariant boundary state of minimal models with unique $\mathbb{Z}_{2}$ symmetry. Since this $\mathbb{Z}_{2}$ symmetry is anomaly free~\cite{PhysRevLett.126.195701}, we expect there are always invariant boundary states.
 
 	For ($A_{m-1}$,$A_{m}$) models, the Cardy state ($a$,$b$) is labeled by points of its Dynkin graph. When $m$ is odd, the $\mathbb{Z}_{2}$ symmetry maps the ($a$,$b$) to ($a$,$m+1-b$). Thus the invariant boundary state is ($a$,$\frac{m+1}{2}$). On the other hand, when $m$ is even, the $\mathbb{Z}_{2}$ symmetry maps the ($a$,$b$) to ($m-a$,$b$). And the invariant boundary state is ($\frac{m}{2}$,$b$).
 	
 	For ($A_{4l}$, $D_{2l+2}$) models, the  $\mathbb{Z}_{2}$ symmetry exchanges two Cardy states if they correspond the same point of $A$ graph and two endpoints of $D$ graph. Thus the other Cardy state is invariant under $\mathbb{Z}_2$ symmetry. Since the action of $\mathbb{Z}_{2}$ symmetry of the  ($A_{4l+2}$,$D_{2l+2}$), ($A_{4l+2}$, $D_{2l+3}$) and ($A_{4l+4}$,$D_{2l+3}$) models,  is same as above, the calculation for the invariant boundary state is also similar.
 	
 	For ($A_{10}$, $E_{6}$) and ($A_{11}$,$E_6$) models,  the $\mathbb{Z}_2$ symmetry is the reflection symmetry of $E_6$ graph. Thus the invariant boundary states correspond to nodes of reflection axis in the $E_6$ graph.
 	
 	\subsection{Invariant boundary states of $S_3$ symmetry}
 	Now we consider the $S_3$ symmetry of ($A_{4}$,$D_4$) (the critical 3-state Potts model) and ($A_{6}$,$D_4$) (the tricritical 3-state Potts model).
 	
 In these two models, the $S_3$ symmetry is the permutation of three outside nodes of the $D_4$ graph. Thus the invariant boundary states correspond to the only one inside node of the $D_4$ graph which implies the $S_3$ symmetry should be anomaly free.
 
 For example, there are eight boundary states for ($A_{4}$,$D_4$) model. The first three states $|A\rangle$, $|B\rangle$, $|C\rangle$
  describe fixed boundary conditions where the spin on the boundary takes one of three possible values. The mixed boundary states $|AB\rangle$,  $|BC\rangle$, $|AC\rangle$ describe boundary conditions where the spin on the boundary can take on two values independently. These six boundary states correspond to six outside nodes of ($A_4$,$D_4$) graph. The remaining two boundary states $|ABC\rangle$, $|N\rangle$ correspond to free boundary conditions and are invariant under the $S_3$ symmetry .
 	
 	Actually, the $\mathbb{Z}_3$ subgroup is $\mathbb{Z}_3$ rotation of 3-state Potts model on the lattice, thus it is anomaly-free as gauging it results in the same theory of the original one~\cite{kikuchi2019twodimensional}. This is nothing but the Kramers-Wannier duality of the 3-state Potts model.

	\section{T-duality symmetry of $\mathrm{SU(2)}_{1}$ WZW model}\label{sec 6}
	In the previous sections, we consider two kinds of symmetries that act on the zero modes of the primary fields of CFTs. The construction of symmetric Cardy states implies that the potential can gap a CFT without spontaneous symmetry breaking. The potentials appearing in the compact boson CFTs are all Haldane gapping potentials. A natural question to be asked is whether symmetric Cardy states can imply the result beyond the usual gapping potentials. 
	
	In this section, we will discuss a related example: the $\mathrm{SU}(2)_{1}$ WZW model with T-duality symmetry.

	\subsection{T-duality symmetry of $\mathrm{SU(2)}_{1}$  WZW model}
	For a free boson CFT with a compacting radius, there is a T-duality  relating theories of radius $2/R$ and $R$. Therefore, only on the self-dual radius $R=\sqrt{2}$ which is exactly  $\mathrm{SU}(2)_{1}$ WZW model, this duality becomes a symmetry which exchanges of two fields 
	\begin{eqnarray}
	T\phi T^{-1}=\theta,\quad
	T\theta T^{-1}=\phi.
	\end{eqnarray}
	This T-duality symmetry not only acts on the zero modes of fields, but also changes every mode in the expansion. In the chiral representation, it acts as follows:
	\begin{eqnarray}
	T\varphi_{0,R} T^{-1}=-\varphi_{0,R},\quad
	T a_{r,R} T^{-1}=-a_{r,R}.
	\end{eqnarray}
	\subsection{Anomaly of T-duality symmetry and center symmetry}
	
	For our interest, we first consider Cardy states invariant under both T-duality symmetry and center symmetry $h$. The $\mathrm{SU}(2)_{1}$ WZW model has $\mathrm{SU}(2)_{L}$$\times$ $\mathrm{SU}(2)_{R}$ transformation and the generators can be written in terms of boson field:
	\begin{eqnarray}
	J^{z}(z)&=&\partial_{z}(\phi+\theta)\ ,\nonumber\\
	J^{+}(z)&=&\exp\left[i\frac{\sqrt{2}}{2}(\phi+\theta)\right],\nonumber\\
	J^{-}(z)&=&\exp\left[-i\frac{\sqrt{2}}{2}(\phi+\theta)\right],\nonumber\\
	\bar{J}^{z}(\bar{z})&=&\partial_{\bar{z}}(-\phi+\theta),\nonumber\\
	\bar{J}^{+}(\bar{z})&=&\exp\left[i\frac{\sqrt{2}}{2}(-\phi+\theta)\right],\nonumber\\
	\bar{J}^{-}({z})&=&\exp\left[-i\frac{\sqrt{2}}{2}(-\phi+\theta)\right].
	\end{eqnarray}
	In this representation, the center symmetry is the minus identity operator $h=-\mathbb{I}$ and the T-duality symmetry acts as $\pi$ rotation along the $x$-axis on the anti-holomorphic boson:
	\begin{eqnarray}
	&&TJ^{z}T^{-1}=J^{z},\nonumber\\
	&& TJ^{\pm}T^{-1}=J^{\pm},\nonumber\\
	&&T\bar{J}^{z}T^{-1}=-\bar{J}^{z},\nonumber\\
	&&T\bar{J}^{\pm}T^{-1}=\bar{J}^{\mp}. 
	\end{eqnarray}
	
	Since the $4\pi$ rotation is the identity in the SU(2) rotation, the T-duality symmetry is a $\mathbb{Z}_{4}$ symmetry in the spin representation. This extension from $\mathbb{Z}_{2}$ to $\mathbb{Z}_{4}$ is because of the center symmetry acting like minus identity \cite{harvey2018uplifting}. In the next section, we will move to the case with only T-duality symmetry where it is a  $\mathbb{Z}_{2}$ symmetry.
	
	Because  $T^{2}$=$h$  after extension, we only need Cardy states invariant under the extended T-duality symmetry:
	\begin{eqnarray}
	T|B\rangle=|B\rangle.
	\end{eqnarray}
	
	To construct such Cardy states, we need to do vector  $\pi/2$ rotation along the $y$-axis:
	\begin{eqnarray}
	&&J^{z}\to J'^{x},\quad
	J'^{x}\to-J'^{z}, \quad
	J'^{y}\to J'^{y},\nonumber\\
	&&\bar{J}^{z}\to\bar{J'}^{x},\quad
	\bar{J}^{x}\to-\bar{J'}^{z},\quad
	\bar{J}^{t}\to\bar{J'}^{y}.
	\end{eqnarray} 
	In the new coordinate,  the T-duality symmetry acts as $\pi$ rotation along the $z$-axis on the antiholomorphic sector:
	\begin{eqnarray}
	&&TJ'^{z}T^{-1}=J'^{z},\nonumber\\
	&& TJ'^{\pm}T^{-1}=J'^{\pm},\nonumber\\
	&&T\bar{J'}^{z}T^{-1}=\bar{J'}^{z},\nonumber\\
	&&T\bar{J'}^{\pm}T^{-1}=-\bar{J'}^{\pm}.
	\end{eqnarray}
	We can also represent these generators in terms of new boson fields $\phi'$ and $\theta'$. Therefore the T-duality acts on the new boson fields as follows:
	\begin{eqnarray}
	&&T\phi' T^{-1}=\phi'+\frac{3\pi\sqrt{2}}{2},\nonumber \\
	&&T\theta' T^{-1}=\theta'+\frac{\pi\sqrt{2}}{2}. 
	\end{eqnarray}
	For one boson theory, there is no symmetric boundary state. However, for 4 copies of $\mathrm{SU}(2)_{1}$ WZW model, we can redefine the boson fields:
	\begin{eqnarray}
	&&\phi''_{1}=\frac{1}{\sqrt{4}}(\phi'_{1}+\phi'_{2}+\phi'_{3}+\phi'_{4}),\nonumber\\
	&&\theta''_{2}=\frac{1}{\sqrt{2}}(\theta'_{1}-\theta'_{2}),\nonumber\\
	&&\theta''_{3}=\frac{1}{\sqrt{2}}(\theta'_{2}-\theta'_{3}),\nonumber\\
	&&\theta''_{4}=\frac{1}{\sqrt{2}}(\theta'_{3}-\theta'_{4}).
	\end{eqnarray}
	The symmetric Cardy state takes the form
	\begin{eqnarray}\label{T duality}
	|B\rangle=\otimes^{4}_{i=1}\left(\frac{1}{\mathscr{N_{i}}}\sum_{v_{i}}|v_{i}\rangle\rangle\right),
	\end{eqnarray}
	where the $v_{i}$ is Ishibashi state for the boson $\phi''_{i}$. The first boson takes the Dirichlet boundary state and the others take Neumann boundary states.
	
	The corresponding gapping potential can be given from the Cardy state
	\begin{eqnarray}
	H'&&=U\lbrack\cos(\frac{\sqrt{2}}{2}(\phi'_{1}+\phi'_{2}+\phi'_{3}+\phi'_{4}))+\cos(\frac{\sqrt{2}}{2}(\theta'_{1}-\theta'_{2}))\nonumber\\
	&&+\cos(\frac{\sqrt{2}}{2}(\theta'_{2}-\theta'_{3}))+\cos(\frac{\sqrt{2}}{2}(\theta'_{3}-\theta'_{4}))\rbrack.
	\end{eqnarray}
	After vector $-\pi/2$ rotation along the $y$-axis, we get the invariant gapping potential under original symmetry:
	\begin{eqnarray}
	H'&&=U\sum^{4}_{j=1}\cos(\frac{\sqrt{2}}{2}\phi_{j})\cos(\frac{\sqrt{2}}{2}\phi_{j+1})\nonumber\\&&+U\sum^{4}_{j=1}\cos(\frac{\sqrt{2}}{2}\theta_{j})\cos(\frac{\sqrt{2}}{2}\theta_{j+1})\nonumber\\&&+U  \text{Re}\prod^{4}_{j=1}\left[\cos(\frac{\sqrt{2}}{2}\phi_{j})+i\cos(\frac{\sqrt{2}}{2}\theta_{j})\right].
	\end{eqnarray}
	Where Re means taking the real part of the third term.
	
	This potential can also be realized as a SU(2) spin ladder model:
	\begin{eqnarray}
	H'&&=U(\lambda_{2})^{2}\sum^{4}_{j=1}\sum_{k} S^{z}_{j,k}S^{z}_{j,k+1}+S^{y}_{j,k}S^{y}_{j,k+1}\nonumber\\&&+U(\lambda_{1})^{4}\sum_{k} \text{Re}\prod_{j=1}^{4}(S_{j,k}\cdot S_{j+1,k}+iS^{x}_{j,k}),
	\end{eqnarray}
	where $\lambda_{1}$ and $\lambda_{2}$ are nonuniversal constants.
	
This potential satisfies the translation symmetry which
is reduced to the center symmetry in the low energy.
However, it remains unclear how to realize the T-duality
symmetry in the lattice model, so we conclude that the
mixed anomaly between the T-duality and center symmetry
is at least stable in a perturbative manner around
this critical point in the field theory. For the perturbation
on the lattice, it is left for future research.

	\subsection{Pure anomaly of T duality symmetry}
	Now we can consider the case of $\mathrm{SU}(2)_{1}$ WZW model with only T-duality symmetry. Since the corresponding SPT phase is trivial \cite{heinrich2018criteria}, we expect the boson CFT can be trivially gapped with persevering T-duality symmetry and has invariant Cardy states.
	
	When there is no $\mathbb{Z}_{2}$ center symmetry, we do not need to see it as a subgroup of $\mathrm{SU}(2)_{R}$.
	To construct the symmetric boundary state, we also apply the vector SU(2) transformation. The fundamental operator is the vertex operator $\exp(\pm i\theta/\sqrt{2})$ and $\exp(\pm i\phi/\sqrt{2})$. This is equivalent to cosine terms  $\cos(\theta/\sqrt{2})$ and  $\cos(\phi/\sqrt{2})$ and similar sine terms. Under  vector $\pi/2$ rotation along $y$-axis, these  terms transform as follows:
	\begin{eqnarray}
	&&\cos(\frac{\sqrt{2}}{2}\theta')=-\sin(\frac{\sqrt{2}}{2}\phi),\nonumber\\
	&&\sin(\frac{\sqrt{2}}{2}\theta')=\sin(\frac{\sqrt{2}}{2}\theta),\nonumber\\
	&&\cos(\frac{\sqrt{2}}{2}\phi')=\cos(\frac{\sqrt{2}}{2}\phi),\nonumber\\
	&&\sin(\frac{\sqrt{2}}{2}\phi')=\cos(\frac{\sqrt{2}}{2}\theta).
	\end{eqnarray}
	In this new coordinate, the T-duality symmetry acts like:
	\begin{eqnarray}
	&&T\cos(\frac{\sqrt{2}}{2}\theta')T^{-1}=-\sin(\frac{\sqrt{2}}{2}\theta'), \nonumber\\
	&&T\sin(\frac{\sqrt{2}}{2}\theta')T^{-1}=-\cos(\frac{\sqrt{2}}{2}\theta'),\nonumber\\
	&&T\cos(\frac{\sqrt{2}}{2}\phi')T^{-1}=\sin(\frac{\sqrt{2}}{2}\phi'),\nonumber\\
	&&T\sin(\frac{\sqrt{2}}{2}\phi') T^{-1}=\cos(\frac{\sqrt{2}}{2}\phi'),
	\end{eqnarray}
	which implies its action on $\theta'$ and $\phi'$:
	\begin{eqnarray}
	\phi'\to\frac{\pi\sqrt{2}}{2}-\phi',\quad
	\theta'\to\frac{3\sqrt{2}\pi}{2}-\theta'.
	\end{eqnarray}
	We can find a symmetric Cardy state:
	\begin{eqnarray}
	|B\rangle=\sum_{v\in\frac{\sqrt{2}}{2}\mathbb{Z}}e^{-i\frac{\sqrt{2}\pi}{4}v}|v\rangle\rangle_{D}.
	\end{eqnarray}
	And	the gapping potential is given by:
	\begin{eqnarray}
	H'=\sqrt{2}U \cos(\frac{\sqrt{2}}{2}\phi'-\frac{\pi}{4}).\label{con:interaction}
	\end{eqnarray}
	
	After rotating back, the corresponding gapping potential is given by:
	\begin{eqnarray}
	H'=U\left[\cos(\frac{\sqrt{2}}{2}\theta)+\cos(\frac{\sqrt{2}}{2}\phi)\right].\label{con:interaction1}
	\end{eqnarray}
	On the lattice, it can be realized as a spin 1/2 antiferromagnetic Heisenberg chain in a staggered magnetic field along the $x$-direction \cite{lecheminant2002criticality}:
	\begin{eqnarray}
	H&&=H_{0}+H'\nonumber\\&&=J\sum_{i}S_{i}\cdot S_{i+1}+U\lambda_{1}\sum_{i}(-1)^{i}S_{i}\cdot S_{i+1}\nonumber\\&&+U\lambda_{2}\sum_{i}(-1)^{i}S^{x}_{i},
	\end{eqnarray}
	where $\lambda_{1}$ and $\lambda_{2}$ are nonuniversal constants.
	
	Since the gapping potential~(\ref{con:interaction}) can condense the boson field and have a unique ground state, the gapping potential~(\ref{con:interaction1}) will also gap the compact boson CFT with a unique ground state after rotating back. As a result, the corresponding ground state on the lattice should be unique.
	
	
	On the lattice, the T-duality symmetry is not an exact symmetry, but can be an emergent symmetry in the low-energy limit. It would be interesting to study the consequence of the emergent T-duality on the lattice model.

	\section{Conclusion and discussion}
	In this work, we discuss the relationship between global anomaly and boundary states of CFTs from the perspective of massive deformation. That is if a 1+1d CFT with symmetry $G$ is anomaly-free,  there will be symmetric boundary conditions where the partition function of BCFT converges to one when the length goes to zero. For the closed string,  symmetric boundary conditions imply the existence of symmetric  boundary states.

	Then we apply our approach to several examples and show the anomaly-free condition derived from the existence of symmetric boundary conditions is consistent with the result from the 't Hooft anomaly-free condition obtained directly from the bulk CFT. In the multicomponent U(1) boson theory, the existence of symmetric boundary conditions can imply the symmetric gapping potentials which belong to Haldane null vectors. We also show this relationship between global anomaly and boundary states can be generalized to  WZW models. As the last example, for   $\mathrm{SU}(2)_{1}$ WZW model with the T-duality symmetry, we use our approach to find the symmetric gapping potentials which are beyond the Haldane null vectors. This result coincides with the fact that there is no anomaly for the T-duality symmetry.
	
	We should note there is a slightly different context discussing the relation between BCFT and 1+1d SPT phases or anomalous 0+1d critical theory \cite{cho2017relationship}. In this reference, it is found that boundary state $|B\rangle_{h}$ in the sector twisted by $h\in G$ can defect the anomalous
phase $\epsilon(g|h)$ which is related to the cocycle in $H^{2}$
($G$, U(1)) :
  \begin{eqnarray}
  g|B\rangle_{h}=\epsilon_{B}(g|h)|B\rangle_{h}.
  \end{eqnarray}
More precisely, this correspondence detects the action of symmetry operation on the twisted boundary condition or twisted boundary state. However, in our approach, we are looking for boundary states in the untwisted sector which remain invariant under the global symmetry $G$.

	As an outlook, it is interesting to consider the relationship between boundary conditions and 't Hooft anomalies in the fermionic CFTs. Here symmetry should include global symmetry $G$ and the fermionic parity symmetry $(-1)^{F}$. The key to solving this problem is searching for the boundary states of a fermionic minimal model. For the bulk fermionic CFTs, they can be constructed from A-type bosonic CFTs~\cite{PhysRevLett.126.195701} and two exceptional CFTs~\cite{Kulp:2020iet} attached with a Kitaev chain after $\mathbb{Z}_{2}$ orbifolding~\cite{Karch:2019aa,Yao2019BosonizationWA} with a parafermionic generalization~\cite{Yao:2021aa}. 
On the lattice model, this transformation is called the Jordan-Wigner transformation and, more generally, the Fradkin-Kadanoff transformation~\cite{Fradkin:1980aa}. 
Naturally, there is a correspondence between boundary states of bosonic CFTs and fermionic CFTs \cite{fukusumi2021fermionization,Ebisu:2021aa}. This correspondence can help us find symmetric boundary states of fermionic CFTs from that of related bosonic CFTs.  Besides, it is quite energizing to apply our BCFT scheme to the intrinsically gapless topological phase in Ref.~\onlinecite{PhysRevB.104.075132,li2022symmetry,Li:2023knf}  where the low energy symmetry is an anomalous $\mathbb{Z}_2$ group but the entire symmetry is a nonanomalous $\mathbb{Z}_4$ group. More precisely, the degree of freedom charged under normal $\mathbb{Z}_2$ symmetry is gapped and $\mathbb{Z}_4$ symmetry does not act on the low energy theory faithfully.  Hence there are no $\mathbb{Z}_4$ symmetric boundary conditions for the low energy theory.  However, one can add a symmetric perturbation to bring the energy of the gapped degree of freedom down.  If further increasing the perturbation reopens the gap, one can eventually arrive at a symmetric gapped ground state and symmetric boundary condition and the non-faithful representation of $\mathbb{Z}_4$ group becomes a faithful one \cite{li2022symmetry}. Thus we conclude that a symmetric boundary condition or gapped ground state exists under a symmetric perturbation if we consider the entire theory of the intrinsically gapless topological phase.

Our analyses in this paper rely heavily on well-developed techniques on CFT in $1+1$ dimensions. 
However, the ``gapping potential argument'' should be applicable to higher-dimensional CFTs as well, and
we also expect universal relationships between the boundary conditions and the anomalies of CFTs
 in higher dimensions. Indeed, this kind of result has been studied in topological quantum field theories by considering fusion algebra of symmetry defects in Ref.~\onlinecite{Thorngren:2020yht}, where anomalous symmetries must be  spontaneously or explicitly broken on the boundary. It is interesting to explore such result directly from the perspective of CFTs in future research.

\section*{Acknowledgements}
L.~L. sincerely thanks Shu-heng Shao and Yunqin Zheng for helpful discussion on the intrinsically gapless topological phase. L.~L. is supported by Global Science Graduate Course (GSGC) program at the
University of Tokyo. 
C.-T.~H. is supported by the Yushan (Young) Scholar Program of the Ministry of Education in Taiwan under grant NTU-111VV016.
Y.~Y. thanks the sponsorship from Yangyang Development Fund and Xiaomi Young Talents Program.
This work was supported in part
by MEXT/JSPS KAKENHI Grants Nos. JP17H06462, JP19H01808, and JP23H01094, JP23K25791, and by JST CREST Grant No. JPMJCR19T2.
	\\

	\bibliography{bib}

	\clearpage
	
	\appendix
	\begin{widetext}
 \section{Haldane gapping potential}
In this appendix, we review some basic knowledge on Haldane gapping potential, which is a universal class of perturbations which can gap the free boson CFTs in 1+1 dimensions\cite{PhysRevLett.74.2090,wang2013non,wang2022symmetric,lu2012theory}.

For $N$ component compact boson CFTs, the Haldane gapping potential is defined as follows:
\begin{eqnarray}
H'=U\sum_{i=1}^{N}\int_{-\infty}^{\infty}dx \cos(\Lambda^{T}_{i}\Phi/R-\alpha_{i}).
\end{eqnarray}
Here $\Phi$=($\phi_{1}$,$\theta_{1}$,....,$\phi_{N}$,$\theta_{N}$) and each $\Lambda_i$ is a 2$N$ dimensional linearly independent vector whose elements are integer. $\alpha_{1}$...$\alpha_{N}$ are arbitrary phases. The above perturbation $H'$ can gap the boson CFT when it satisfies the following sufficient conditions \cite{PhysRevLett.74.2090}:
\begin{eqnarray}\label{eq:condition1}
\Lambda^{T}_{i}K\Lambda_{j}=0, ~~\forall i,j.
\end{eqnarray}
Here $K=\oplus^{N}_{i=1}\sigma^{i}_{x}$ and such vectors $\Lambda_{i}$ are also be called ``null-vectors".

Now we can further assume the perturbation is invariant under a global symmetry group $G$:
\begin{eqnarray}
g^{-1}H'g=H'\ ,\quad \forall g\in G.
\end{eqnarray}
Although the perturbation is symmetric, the ground state may break the symmetry spontaneously and have nontrivial degeneracy. There is another primitive condition which can forbid this case and ensure the unique gapped ground state. More precisely, the primitive condition means there is no solution to the equation:
\begin{eqnarray}\label{eq:condition2}
a_{1}\Lambda_{1}+...+a_{N}\Lambda_{N}=k\Lambda,
\end{eqnarray}
where $a_{i}$ are integers with no common divisors, $k$ is an integer that is greater than 1, and $\Lambda$ is an integer vector.

 Indeed, the compact boson CFTs can be realized as the edge theories of 2+1d Abelian Chern-Simons (CS) theory \cite{wen1995topological} which can describe the quantum Hall effect.  Such Haldane gapping potential can be also understood as the scattering process of electrons or quasiparticles on the edge between different modes. Ref.~\onlinecite{PhysRevX.3.021009} shows the condition \eqref{eq:condition1} and \eqref{eq:condition2} can also be understood from the gapped boundary conditions of Abelian CS theory. 
\section{Dictionary connecting spin operators and boson fields}
In this appendix, we will review the dictionary connecting spin operators and boson fields, which can be derived by Jordan-Wigner transformation and bosonization. This appendix is mainly based on Ref.~\onlinecite{senechal2004introduction,PhysRevB.93.104425}.

We start with the spin-1/2 XXZ model with microscopic Hamiltonian:
\begin{equation}
    H=\sum_{n}[J(S^x_nS^x_{n+1}+S^y_nS^y_{n+1})+J_zS^z_nS^z_{n+1}].
\end{equation}
Thanks to the Jordan-Wigner transformation:
\begin{equation}
   S^{+}_n=c^{+}_n\exp(i\pi \sum^{n-1}_{m=1}c^{+}_mc_m), S^z_n=c^{+}_nc_n-\frac{1}{2}, \quad c^{+}_{n}c_{m}+c_{m}c^{+}_{n}=\delta_{mn},
\end{equation}
the above Hamiltonian can be mapped to a spinless fermion Hamiltonian:
\begin{equation}
    H=\sum_{n}[-\frac{J}{2}(c^{+}_nc_{n+1}+H.c.)+J_z(c^{+}_nc_n-\frac{1}{2})(c^{+}_{n+1}c_{n+1}-\frac{1}{2})].
\end{equation}
In the low energy, the fermion chain can be described by a Dirac fermion theory with four fermion interactions:
\begin{equation}
    H_{\text{low}}=J\int dx (\psi^{+}_R\partial_x \psi_R-\psi^{+}_L\partial_x \psi_L)+J_z\int dx[2\psi^{+}_L\psi_L\psi^{+}_R\psi_R-(\psi^+_R\psi_L+\psi^+_L\psi_R)^2],
\end{equation}
where we choose lattice spacing being 1.
By the bosonization approach, this model can be mapped to the compact boson CFT with interactions:
\begin{equation}
H_{\text{low}}=\frac{J}{8\pi}\int dx (\partial_{x}\phi)^2+(\partial_{x}\theta)^2-2J_z\cos(\frac{2\phi}{R}),
\end{equation}
with radius 
\begin{equation}
  \phi\equiv \phi+2\pi R,~\theta \equiv \theta+4\pi/R,~  R=\sqrt{2}[1-\frac{1}{\pi}\cos^{-1}(\frac{J_z}{J})]^{\frac{1}{2}},
\end{equation}
where $\theta$ is a dual field of $\phi$, i.e., $\partial_x\theta=\partial_t \phi$. When $|J_z|\le |J|$, the interaction term is irrelevant and thus  the low energy physics of spin-1/2 XXZ chain can be described by the compact boson CFT.

The correspondence between spin operators and the boson field is given by
\begin{equation}
\begin{split}
    S^z(x)\approx\frac{1}{2\pi R}\partial_x \phi+(-1)^{x}\frac{\lambda}{\pi }\sin(\phi/R), \quad S^{+}(x)\approx e^{i\theta R/2}[\frac{\lambda}{\pi a}(-1)^{x}+\sin(\phi/R)],
    \end{split}
\end{equation}
with $x=n$.
It is easy to check the translation and $U(1)_z$ acts on the low energy boson fields as
\begin{equation}
\begin{split}
    &T: \phi\to \phi+\pi R,~ \theta \to \theta+\pi R, \\
    &U(1)_z: \phi \to \phi, ~ \theta\to \theta+\alpha R.
    \end{split}
\end{equation}
		\section{$\mathrm{SU}(N)$ symmetry invariant boundary states of free boson model}
		In this appendix, we will use the affine Lie algebra to show the boundary states (\ref{bstate1})(\ref{bstate2}) are invariant under vector $\mathrm{SU}(N)$  spin rotation.
		
		Firstly, we prove the result for the case of SU(2). For simplicity, the boundary state (\ref{bstate1}) can be written in the language of chiral operators:
		\begin{eqnarray}
		|B\rangle&&=\frac{e^{\sum_{r\in \mathbb{N_{+}}}\frac{-1}{r}(a^{1}_{-r,L}a^{1}_{-r,R}+a^{2}_{-r,L}a^{2}_{-r,R}})}{\mathscr{N}_{N}\mathscr{N}_{D}}\sum_{m,n}(e^{in\frac{\sqrt{2}}{2}(\theta^{0}_{1}-\theta^{0}_{2})}-e^{im\frac{\sqrt{2}}{2}(\phi^{0}_{1}+\phi^{0}_{2})})e^{-\frac{1}{r}a^{1}_{-r,L}a^{1}_{-r,R}}e^{-\frac{1}{r}a^{2}_{-r,L}a^{2}_{-r,R}}|0\rangle\nonumber\\
		&&=\frac{e^{\sum_{r\in \mathbb{N_{+}}}\frac{-1}{r}(a^{1}_{-r,L}a^{1}_{-r,R}+a^{2}_{-r,L}a^{2}_{-r,R}})}{\mathscr{N}_{N}\mathscr{N}_{D}}\sum_{m,n}(e^{in\frac{\sqrt{2}}{2}(\varphi^{0}_{1,L}-\varphi^{0}_{1,R}-\varphi^{0}_{2,L}+\varphi^{0}_{2,R})}-e^{im\frac{\sqrt{2}}{2}(\varphi^{0}_{1,L}+\varphi^{0}_{1,R}+\varphi^{0}_{2,L}+\varphi^{0}_{2,R})})|0\rangle,\nonumber\\
		&&~
		\end{eqnarray}
		where $\varphi^{0}_{L}$ /$\varphi^{0}_{R}$ is zero mode of primary fields  $\varphi_{L}(\omega=0)$/$\varphi_{R}(\bar{\omega}=0)$.
		
		The affine SU(2) generators are given by:
		\begin{eqnarray}
		&&H=i\partial\varphi_{L},\quad
		E^{\pm}=e^{\pm i\sqrt{2}\varphi_{L}},\nonumber\\
		&&\bar{H}=-i\bar{\partial}\varphi_{R},\quad\bar{E}^{\pm}=e^{\mp i\sqrt{2}\varphi_{R}}.
		\end{eqnarray}
		The OPE between affine SU(2) generators and primary fields are:
		\begin{eqnarray}\label{ope}
		&&H(z) e^{\pm i\frac{1}{\sqrt{2}}\varphi^{0}_{L}}=\frac{\pm e^{\pm\frac{1}{\sqrt{2}}\varphi^{0}_{L}}}{\sqrt{2}z},\quad
		E^{\pm}(z)e^{\mp i\frac{1}{\sqrt{2}}\varphi^{0}_{L} }=\frac{e^{\pm i\frac{1}{\sqrt{2}}\varphi^{0}_{L} }}{z},\nonumber\\
		&&\bar{H}(\bar{z}) e^{\pm i\frac{1}{\sqrt{2}}\varphi^{0}_{R}}=\frac{\mp e^{\pm\frac{1}{\sqrt{2}}\varphi^{0}_{R}}}{\sqrt{2}\bar{z}},\quad
		\bar{E}^{\pm}(\bar{z})e^{\pm i\frac{1}{\sqrt{2}}\varphi^{0}_{R} }=\frac{e^{\mp i\frac{1}{\sqrt{2}}\varphi^{0}_{R} }}{\bar{z}}.
		\end{eqnarray}
		
		Since the generators of SU(2) rotation are zero modes of affine SU(2) generators, they satisfy the following OPE with primary fields:
		\begin{eqnarray}
		&&H^{0} e^{\pm i\frac{1}{\sqrt{2}}\varphi^{0}_{L}}=\pm e^{\pm\frac{1}{\sqrt{2}}\varphi^{0}_{L}},\quad
		E^{0,\pm}e^{\mp i\frac{1}{\sqrt{2}}\varphi^{0}_{L} }=e^{\pm i\frac{1}{\sqrt{2}}\varphi^{0}_{L} },\nonumber\\
		&&\bar{H}^{0} e^{\pm i\frac{1}{\sqrt{2}}\varphi^{0}_{R}}=\mp e^{\pm\frac{1}{\sqrt{2}}\varphi^{0}_{R}},\quad
		\bar{E}^{0,\pm}e^{\pm i\frac{1}{\sqrt{2}}\varphi^{0}_{R} }=e^{\mp i\frac{1}{\sqrt{2}}\varphi^{0}_{R} }.
		\end{eqnarray}
		
		Since the other OPEs are zero, the SU(2) rotation only acts on the following parts of the boundary state:
		\begin{eqnarray}\label{eq:boundary state}
		&&\lbrack e^{i\frac{\sqrt{2}}{2}(\varphi^{0}_{1,L}-\varphi^{0}_{1,R}-\varphi^{0}_{2,L}+\varphi^{0}_{2,R})}-e^{i\frac{\sqrt{2}}{2}(\varphi^{0}_{1,L}+\varphi^{0}_{1,R}+\varphi^{0}_{2,L}+\varphi^{0}_{2,R})}\nonumber\\&&+e^{-i\frac{\sqrt{2}}{2}(\varphi^{0}_{1,L}-\varphi^{0}_{1,R}-\varphi^{0}_{2,L}+\varphi^{0}_{2,R})}-e^{-i\frac{\sqrt{2}}{2}(\varphi^{0}_{1,L}+\varphi^{0}_{1,R}+\varphi^{0}_{2,L}+\varphi^{0}_{2,R})}\rbrack |0\rangle.
		\end{eqnarray}
		
		For the vector SU(2) rotation, the generators is $S^{z}=H^{0}+\bar{H}^{0}$, $S^{\pm}=E^{0,\pm}+\bar{E}^{0,\pm}$.  In this representation, the SU(2) generators are the sum of four spin-1/2 generators. The following states can be written as the spin-1/2 since primary fields transform like a spinor:
\begin{equation}
    \begin{split}
      \exp(i\varphi^{0}_{L}/\sqrt{2})|0\rangle= |\uparrow\rangle,\quad \exp(-i\varphi^{0}_{L}/\sqrt{2})|0\rangle= |\downarrow\rangle, \\
      \exp(-i\varphi^{0}_{R}/\sqrt{2})|0\rangle= |\uparrow\rangle,\quad \exp(i\varphi^{0}_{R}/\sqrt{2})|0\rangle= |\downarrow\rangle.
    \end{split}
\end{equation}
where the spin up and down is determined by the sign of eigenvalue of $H^0$ or $\bar{H}^0$.
  Then the boundary state \eqref{eq:boundary state} can be rewritten as 
		\begin{eqnarray}
		&&|\uparrow_{1,L}\uparrow_{1,R}\downarrow_{2,L}\downarrow_{2,R}\rangle\rangle-|\uparrow_{1,L}\downarrow_{1,R}\uparrow_{2,L}\downarrow_{2,R}\rangle\rangle+|\downarrow_{1,L}\downarrow_{1,R}\uparrow_{2,L}\uparrow_{2,R}\rangle\rangle-|\downarrow_{1,L}\uparrow_{1,R}\downarrow_{2,L}\uparrow_{2,R}\rangle\rangle\nonumber\\
		&&=(|\uparrow_{1,L}\downarrow_{2,R}\rangle\rangle-|\downarrow_{1,L}\uparrow_{2,R}\rangle\rangle)\otimes(|\uparrow_{1,R}\downarrow_{2,L}\rangle\rangle-|\downarrow_{1,R}\uparrow_{2,L}\rangle\rangle).
		\end{eqnarray}
		 It is easy to see the boundary state is invariant under vector SU(2) transformation since it is the direct product of two spin singlet states.
		
		For general vector $\mathrm{SU}(N)$  transformation, the Lie algebra can be composed into $N(N-1)/2$ SU(2) Lie subalgebras. For each SU(2) Lie subalgebras, we can do similar calculations to show the eigenvalue of SU(2) generators is zero. So the boundary state is invariant under vector $\mathrm{SU}(N)$  spin rotation.  
		
		\section{Invariant boundary state of vector rotation symmetry and center symmetry}\label{app_B}
		In this appendix, we will give a detailed calculation on invariant boundary states of vector rotation symmetry and center symmetry in the WZW model. 
		
		The Ishibashi state is defined as follows:
		\begin{eqnarray}
		|\lambda\rangle\rangle=\sum_{m}|\phi_{i}^{\lambda},m\rangle\otimes U|\bar{\phi}^{\lambda}_{i},\bar{m}\rangle.
		\end{eqnarray}
		
		To show the Ishibashi state is invariant under the vector rotation symmetry, we only need to show the action of generators on such states is zero. Since the holomorphic and antiholomorphic sector is tensor product, the action of a generator is a summation of action on each part : $S^{a}=J^{a}_{0}+\bar{J}^{a}_{0}$.  The vacuum states of each sector form different irreducible representations of Lie algebra labeling by $\lambda$:
		\begin{eqnarray}
		J^{a}_{0}|\phi^{\lambda}_{i},m\rangle=\sum_{j} T^{\lambda,a}_{ij}|\phi^{\lambda}_{j},m\rangle,\nonumber\\
		\bar{J}^{a}_{0}|\bar{\phi}^{\lambda}_{i},m\rangle=\sum_{j} T^{\lambda,a}_{ij}|\bar{\phi}^{\lambda}_{j},m\rangle.
		\end{eqnarray}
		
		Here $U$ operator is an anti-unitary operator and anti-commutes with Lie algebra:
		\begin{eqnarray}
		U^{-1} \bar{J}^{a}_{n} U=-(\bar{J}^{a}_{-n})^{+}=-(\bar{J}^{a}_{n}).
		\end{eqnarray}
		
		Now we can show  the action of generators on each vacuum in the summation is zero:
		\begin{eqnarray}
		&&\sum_{m}\langle \phi^{\lambda}_{j},n |\otimes \langle U\bar{\phi_{k}}^{\lambda},n'| S^{a}|\phi^{\lambda}_{i},m\rangle\otimes U|\bar{\phi_{i}}^{\lambda},m\rangle\nonumber\\&&=\sum_{m}\delta_{ik}\delta_{n'm}\langle \phi^{\lambda}_{j},n |J^{a}_{0}|\phi^{\lambda}_{i},m\rangle+\delta_{ij}\delta_{nm}\langle U\bar{\phi}^{\lambda}_{k},n' |\bar{J}^{a}_{0}U|\bar{\phi}^{\lambda}_{i},m\rangle\nonumber\\
		&&=\delta_{ik}\delta_{ij}(\langle \phi^{\lambda}_{i},n |J^{a}_{0}|\phi^{\lambda}_{i},n'\rangle+\langle U\bar{\phi}^{\lambda}_{i},n' |U^{+}\bar{J}^{a}_{0}U|\bar{\phi}^{\lambda}_{i},n\rangle)\nonumber\\&&=\delta_{ik}\delta_{jk}(\langle \phi^{\lambda}_{i},n |J^{a}_{0}|\phi^{\lambda}_{i},n'\rangle-\langle \bar{\phi}^{\lambda}_{i},n' |\bar{J}^{a}_{0}|\bar{\phi}^{\lambda}_{i},n\rangle^{*})\nonumber\\&&=0.
		\end{eqnarray}
		
		Thus the Cardy state is also invariant under vector rotation symmetry since it is a linear summation of Ishibashi states. The condition for invariant Cardy states is now:
		\begin{eqnarray}
		h|B\rangle=|B\rangle.
		\end{eqnarray}
		\subsection{$B_{N}$ affine Lie algebra}
		For $B_{N}$ type Lie algebra,
		the action of the outer automorphism  is 
		\begin{eqnarray}
		&&A:\lbrack\lambda_{0};\lambda_{1},\cdots,\lambda_{N-1},\lambda_{N}\rbrack\to\lbrack\lambda_{1};\lambda_{0},\cdots,\lambda_{N-1},\lambda_{N}\rbrack.
		\end{eqnarray}
		Thus the affine Dynkin labels of an invariant boundary state satisfy
		
		\begin{eqnarray}
		\lambda_{0}=\lambda_{1}.
		\end{eqnarray} 
		
		Since the comarks of $B_{N}$ are:
		\begin{eqnarray}
		(a^{\vee}_{0};a^{\vee}_{1},\cdots,a^{\vee}_{N})=(1;1,2,\cdots,2,2,1),
		\end{eqnarray}
		We can obtain the level of $k$

		\begin{equation}
		k=\sum^N_{i=0}a^{\vee}_{i}\lambda_i=2(\lambda_{1}+\cdots+\lambda_{N-1})+\lambda_{N}. 
		\end{equation}
		
		Thus we can always find a Cardy state invariant under the vector $\mathrm{SO}(2l+1)$ symmetry and center symmetry.
		\subsection{$D_{2l}$ affine Lie algebra}
		For $D_{2l}$ type Lie algebra, there are two outer automorphisms:
	\begin{eqnarray}
	&& A_{1}:\lbrack\lambda_{0};\lambda_{1},\lambda_{2},\cdots,\lambda_{2l-1},\lambda_{2l}\rbrack\to\lbrack\lambda_{1};\lambda_{0},\lambda_{2},\cdots,\lambda_{2l},\lambda_{2l-1}\rbrack,\nonumber\\
	&& A_{2}:\lbrack\lambda_{0};\lambda_{1},\lambda_{2},\cdots,\lambda_{2l-2},\lambda_{2l-1},\lambda_{2l}\rbrack\to\lbrack\lambda_{2l};\lambda_{2l-1},\lambda_{2l-2},\cdots,\lambda_{2},\lambda_{1},\lambda_{0}\rbrack.
	\end{eqnarray}
		So the affine Dynkin labels of an invariant boundary state satisfy
	\begin{eqnarray}
	\lambda_{1}&=&\lambda_{0}=\lambda_{2l}=\lambda_{2l-1},\nonumber\\
	\lambda_{j}&=&\lambda_{2l-j}\ ( j\ge 2).
	\end{eqnarray}
		
		The comarks of $D_{N}$ are:
		\begin{eqnarray}
		(a^{\vee}_{0};a^{\vee}_{1},\cdots,a^{\vee}_{N})=(1;1,2,\cdots,2,1,1).
		\end{eqnarray}       As a result, for the existence of an invariant boundary state, $k$ satisfies
		\begin{eqnarray}
		k=4\lambda_{0}+4(\lambda_{2}\cdots+\lambda_{l-1})+2\lambda_{l}.
		\end{eqnarray}
		
		Therefore, there exists an invariant Cardy state under vector $\mathrm{SO}(4l)$ symmetry and center symmetry only for even $k$.
		\subsection{$C_{N}$ affine Lie algebra}
		For $C_{N}$ type Lie algebra,
		the action of the outer automorphism  is 
		\begin{eqnarray}
		&&A:\lbrack\lambda_{0};\lambda_{1},\cdots,\lambda_{N-1},\lambda_{N}\rbrack\to\lbrack\lambda_{N};\lambda_{N-1},\cdots,\lambda_{1},\lambda_{0}\rbrack.
		\end{eqnarray}
		Thus the affine Dynkin labels of an invariant boundary state satisfy
		
		\begin{eqnarray}
		\lambda_{j}=\lambda_{N-j}.
		\end{eqnarray} 
		
		Since the comarks of $C_{N}$ are:
		\begin{eqnarray}
		(a^{\vee}_{0};a^{\vee}_{1},\cdots,a^{\vee}_{N})=(1;1,\cdots,1),
		\end{eqnarray}
		We can obtain the level of $k$ 
		
		\begin{equation}
		k=
		\begin{cases}
		2(\lambda_{0}+\cdots+\lambda_{\frac{N}{2}-1})+\lambda_{\frac{N}{2}},  & \mbox{if }N\mbox{ is even}; \\  2(\lambda_{0}+\cdots+\lambda_{\frac{N-1}{2}}),  & \mbox{if }N\mbox{ is odd} .
		\end{cases}
		\end{equation}

		Therefore only when $k$ is even or $N$ is even, there will be a Cardy state invariant under the vector $\mathrm{Usp}(N)$ symmetry and center symmetry.
		\subsection{$D_{2l+1}$ affine Lie algebra}
		For $D_{2l+1}$ type Lie algebra,
		the action of the outer automorphism  is 
		\begin{eqnarray}
		&&A:\lbrack\lambda_{0};\lambda_{1},\cdots,\lambda_{2l},\lambda_{2l+1}\rbrack\to\lbrack\lambda_{2l};\lambda_{2l+1},\cdots,\lambda_{1},\lambda_{0}\rbrack.
		\end{eqnarray}

		The affine Dynkin labels of an invariant boundary state satisfy
		
		\begin{eqnarray}
		\lambda_{0}=\lambda_{2l}=\lambda_{1}=\lambda_{2l+1},\lambda_{j}=\lambda_{2l+1-j}\ ( 2\le j\le 2l-1).
		\end{eqnarray} 
		
		Since the comarks of $D_{N}$ are:
		\begin{eqnarray}
		(a^{\vee}_{0};a^{\vee}_{1},\cdots,a^{\vee}_{N})=(1;1,2,\cdots,2,1,1),
		\end{eqnarray}
		We can obtain the level of $k$ 
		
		\begin{equation}
		k=4(\lambda_{0}+\lambda_{2}+\cdots+\lambda_{l}).
		\end{equation}

		Therefore only when $k$ is a multiple of 4, there will be a Cardy state invariant under the vector $\mathrm{SO}(4l+2)$ symmetry and center symmetry.
		
		\subsection{$E_{6}$ affine Lie algebra}
		For $E_{6}$ type Lie algebra,
		the action of the outer automorphism  is 
		\begin{eqnarray}
		&&A:\lbrack\lambda_{0};\lambda_{1},\cdots,\lambda_{6}\rbrack\to\lbrack\lambda_{1};\lambda_{5},\lambda_{4},\lambda_{3},\lambda_{6},\lambda_{0},\lambda_{2}\rbrack.
		\end{eqnarray}
		The affine Dynkin labels of an invariant boundary state satisfy
		
		\begin{eqnarray}
		\lambda_{0}=\lambda_{5}=\lambda_{1},\quad \lambda_{2}=\lambda_{4}=\lambda_{6}.
		\end{eqnarray}
		
		Since the comarks of $E_{6}$ are:
		\begin{eqnarray}
		(a^{\vee}_{0};a^{\vee}_{1},\cdots,a^{\vee}_{6})=(1;1,2,3,2,1,2),
		\end{eqnarray} the level $k$ is given by
		\begin{eqnarray}
		k=\lambda_{0}+\lambda_{1}+2\lambda_{2}+3\lambda_{3}+2\lambda_{4}+\lambda_{5}+2\lambda_{6}=3(\lambda_{0}+\lambda_{3}+2\lambda_{2}).
		\end{eqnarray}
		
		This implies only if $k$ is a multiple of 3, there will be a Cardy state invariant under the vector $E_{6}$ symmetry and center symmetry.
		\subsection{$E_{7}$ affine Lie algebra}
		For $E_{7}$ type Lie algebra,
		the action of the outer automorphism  is 
		\begin{eqnarray}
		&&A:\lbrack\lambda_{0};\lambda_{1},\cdots,\lambda_{7}\rbrack\to\lbrack\lambda_{6};\lambda_{5},\lambda_{4},\lambda_{3},\lambda_{2},\lambda_{1},\lambda_{0},\lambda_{7}\rbrack.
		\end{eqnarray}
		The affine Dynkin labels of an invariant boundary state satisfy
		
		\begin{eqnarray}
		\lambda_{0}=\lambda_{6},\quad \lambda_{1}=\lambda_{5},\quad \lambda_{2}=\lambda_{4}=\lambda_{4}. 
		\end{eqnarray}
		
		Since the comarks of $E_{7}$ are:
		\begin{eqnarray}
		(a^{\vee}_{0};a^{\vee}_{1},\cdots,a^{\vee}_{7})=(1;2,3,4,3,2,1,2),
		\end{eqnarray}
		the level $k$ is given by
		\begin{eqnarray}
		k=2\lambda_{0}+4\lambda_{1}+6\lambda_{2}+4\lambda_{3}+2\lambda_{7}.
		\end{eqnarray}
		
		This implies only if $k$ is even, there will be a Cardy state invariant under the vector $E_{7}$ symmetry and center symmetry.
		
		\section{Invariant boundary state of  vector rotation symmetry and $\mathcal{CP}_{2}$ symmetry}\label{app_c}
		In this appendix, we will give a derivation on the invariant boundary state of vector rotation symmetry and $\mathcal{CP}_{2}$ symmetry in the WZW model. As shown in the appendix \ref{app_B}, all Cardy states are invariant under vector rotation and  spatial reflection symmetry, we only need to search for Cardy states which are invariant under the combination of charge conjugation $C$ and an order 2 element of center symmetry $h_2$. Since the charge conjugation acts as an identity on Cardy states and center symmetry is a $\mathbb{Z}_2$ group for $B_{N}$, $C_{N}$, $D_{2l}$, $E_{7}$ cases, the result is the same as that of appendix \ref{app_B}. Moreover, the center symmetry of $E_{6}$ WZW model is $\mathbb{Z}_3$ group which implies we can't construct $\mathcal{T}_2$ symmetry. Thus  we only need to calculate the $D_{2l+1}$ WZW model where the order 2 element is $A^2$.

		For $D_{2l+1}$ type Lie algebra,
		the action of the $\mathbb{Z}^{Ch_2}_{2}$ transformation  is 
		\begin{eqnarray}
		&&CA^2:\lbrack\lambda_{0};\lambda_{1},\cdots,\lambda_{2l},\lambda_{2l+1}\rbrack\to\lbrack\lambda_{1};\lambda_{0},\cdots,\lambda_{2l},\lambda_{2l+1}\rbrack.
		\end{eqnarray}
The affine Dynkin labels of an invariant boundary state satisfy
		
		\begin{eqnarray}
		\lambda_{0}=\lambda_{1}.
		\end{eqnarray}

		Then when there is an invariant boundary state, $k$ satisfies:
		
		\begin{equation}
		k=2\lambda_{0}+2(\lambda_{2}+\cdots+\lambda_{2l-1})+\lambda_{2l}+\lambda_{2l+1}.
		\end{equation}
		Therefore, there always exists an invariant boundary state for arbitrary $k$.

		\section{Symmetric interaction of $\mathrm{SU}(2)_{2k}$ WZW model}\label{app d}
		The calculation in this appendix is based on an argument \cite{cardy2017bulk} for a perturbed CFT with the Hamiltonian:
		\begin{eqnarray}
		H=H_{CFT}+ \sum_{j}\lambda_{j}\int\Phi^{j}(x) dx.
		\end{eqnarray}  
		Here $\Phi^{j}$ are relevant operators (primary fields). It claims that the smeared boundary state  $e^{-\tau_{a} H}|B_{a}\rangle$ with  the lowest variational energy can be very close to the ground state. This variational energy is given by:
		\begin{eqnarray}
		E_{a}=\frac{\pi c}{96\tau_{a}^{2}}+\sum_{j\ne 0}\lambda_{j}\frac{S^{j}_{a}}{S^{0}_{a}}\left(\frac{S^{0}_{0}}{S^{j}_{0}}\right)^{\frac{1}{2}}\frac{\pi^{\triangle_{j}}}{(2\tau_{a})^{\triangle_{j}}}.
		\end{eqnarray}
		For the $\mathrm{SU}(2)_{2k}$ WZW model, the relevant operator $\lambda\tr(g^{2})$ is the primary field $\lambda \Phi^{1}$. We have $2k+1$ boundary states $|B_{a}\rangle$ for $2a\in \mathbb{Z}$ and $0\le a\le k$. Thus the variational energy of the smeared boundary state  $e^{-\tau_{a} H}|B_{a}\rangle$ is:
		\begin{eqnarray}
		E_{a}=\frac{\pi c}{96\tau_{a}^{2}}+\lambda\frac{S^{1}_{a}}{S^{0}_{a}}\left(\frac{S^{0}_{0}}{S^{1}_{0}}\right)^{\frac{1}{2}}\left(\frac{\pi}{2\tau_{a}}\right)^{\frac{2}{k+1}}.
		\end{eqnarray}
		When	$\lambda>0$, we can vary $\tau_{a}$ and obtain the minimal of $E_{a}$:
		\begin{equation}
		min(E_{a})=
		\begin{cases}
		0,  & \mbox{if }\frac{S^{1}_{a}}{S^{0}_{a}}\ge0; \\  -\lambda b\left|\frac{S^{1}_{a}}{S^{0}_{a}}\right|^{\frac{k+1}{k}},  & \mbox{if }\frac{S^{1}_{a}}{S^{0}_{a}}<0,
		\end{cases}
		\end{equation}
		where $b$ is a positive number only depending on $k$. 
		
		The modular $S$ matrix of $\mathrm{SU}(2)_{2k}$ WZW model are given by:
		\begin{eqnarray}
		S^{j'}_{j}=\sqrt{\frac{1}{k+1}}\sin \frac{\pi (2j+1)(2j'+1)}{2k+2}.
		\end{eqnarray}
		Thus the minimal of $E_{a}$ can be rewritten as:
		\begin{equation}
		min(E_{a})=
		\begin{cases}
		0,  & \mbox{if } a\le\frac{2k-1}{6} \mbox{ or } a\ge \frac{4k+1}{6}; \\  -\lambda b\left|4\cos^{2}(\frac{2a+1}{2k+2}\pi)-1\right|^{\frac{k+1}{k}},  & \mbox{if } \frac{2k-1}{6}\le a\le\frac{4k+1}{6}.
		\end{cases}
		\end{equation}
		It is easy to see when $a=k/2$, the minimal of variational energy is lowest. In the language of Dykin label, this boundary state is $\lbrack k;k\rbrack$ which is the only invariant state under the center symmetry. 
		
		When $\lambda<0$, a similar calculation can show that the variational energy is lowest  when $a=0$ or $a=k$. These two states correspond to the Valence Bond Solid (VBS) phase which breaks translation symmetry.  
		
		We can also show the  massive phase with $\lambda>0$ is the Haldane phase if $k$ is odd and is the trivial phase if $k$ is even. To see this, we consider such massive deformation:
		\[V(x)=\begin{cases}
		
		\lambda \tr(g^{2}),&x<0;\\
		
		0,&0 \leq x < L;\\
		-\lambda \tr(g^{2}),&L \le x.
		
		\end{cases}\]
		The partition function of the middle gapless region is given by
		\begin{eqnarray}
		Z_{\frac{k}{2}0}=Z_{\frac{k}{2}k}=\chi_{\frac{k}{2}}(q)=(k+1)q^{\triangle_{\frac{k}{2}}}+\cdots.
		\end{eqnarray}
		Here $\cdots$ is polynomials of $q$ whose degree is large than $\triangle_{k/2}$. 
		
		Thus when $L$ goes to zero, there will be $k+1$ states on the interface which are equivalent to states of spin-$k/$2. When $k$ is even, the effective spin on the interface is an integer which can be gapped without degeneracy. When $k$ is odd, the effective spin is a half-integer which is always degenerate. We show the lattice version  of the interface for $\mathrm{SU}(2)_{2}$ WZW in the figure \ref{c}. Here dots are the effective spin 1/2 decoupled from the spin 1 on each site. Thus the unpaired spin 1/2 on the interface implies the symmetry protected degeneracy.
		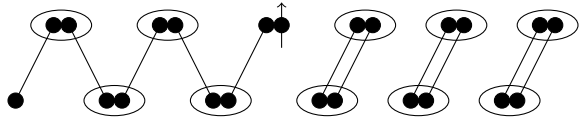
\begin{figure}
			\centering
			\begin{tikzpicture}
			
			\draw (5.1,1) ellipse (0.4 and 0.2);
			\filldraw (5,1) circle (.1);
			\filldraw (5.2,1) circle (.1);
			
			\draw (4.4,2) ellipse (0.4 and 0.2);	
			\filldraw (4.5,2) circle (.1);
			\filldraw (4.3,2) circle (.1);
			
			\filldraw (5.7,2) circle (.1);
			
			\draw (3.7,1) ellipse (0.4 and 0.2);
			\filldraw (3.8,1) circle (.1);
			\filldraw (3.6,1) circle (.1);
			
			\draw (3,2) ellipse (0.4 and 0.2);
			\filldraw (3.1,2) circle (.1);
			\filldraw (2.9,2) circle (.1);
			
			\filldraw (2.4,1) circle (.1);
			
			\draw (5,1)--(4.5,2);
			\draw (5.2,1)--(5.7,2);
			\draw (3.8,1)--(4.3,2);
			\draw (3.6,1)--(3.1,2);
			\draw (2.4,1)--(2.9,2);
			\filldraw (5.9,2) circle (.1);
			\draw[->] (5.9,1.7) --(5.9,2.3);
			\draw (6.5,1) ellipse (0.4 and 0.2);
			\draw (7,2) ellipse (0.4 and 0.2);
			\filldraw (6.4,1) circle (.1);
			\filldraw (6.6,1) circle (.1);
			\filldraw (6.9,2) circle (.1);
			\filldraw (7.1,2) circle (.1);
			\draw (6.4,1)--(6.9,2);
			\draw (6.6,1)--(7.1,2);
			
			\draw (7.7,1) ellipse (0.4 and 0.2);
			\draw (8.2,2) ellipse (0.4 and 0.2);
			\filldraw (7.6,1) circle (.1);
			\filldraw (7.8,1) circle (.1);
			\filldraw (8.1,2) circle (.1);
			\filldraw (8.3,2) circle (.1);
			\draw (7.6,1)--(8.1,2);
			\draw (7.8,1)--(8.3,2);
			
			\draw (8.9,1) ellipse (0.4 and 0.2);
			\draw (9.4,2) ellipse (0.4 and 0.2);
			\filldraw (8.8,1) circle (.1);
			\filldraw (9,1) circle (.1);
			\filldraw (9.3,2) circle (.1);
			\filldraw (9.5,2) circle (.1);
			\draw (8.8,1)--(9.3,2);
			\draw (9,1)--(9.5,2);
			
			\end{tikzpicture}
			\centering
			\captionof{figure}{\textbf{The lattice version of interface between Haldane phase and VBS phase  of $\mathrm{SU}(2)_{2}$ WZW model.}}
			\label{c}
		\end{figure}

\section{The LSM ingappability of the spin chain with triple-product interactions }\label{LSM}
In this appendix, we will prove the LSM ingappability of the Hamiltonian \eqref{triple-product} with the following symmetry:
\begin{eqnarray}\label{eq:sym}
&&\mathcal{T}_2: T_2=TKU_{T}=TK\prod^{j=L}_{j=1}i\sigma^y_j,\\
&&U(1)_z: U_{\theta}=\exp(i\theta\sum^{L}_{j=1}  \bold{S}^z_{j}) ,\\
&&\mathbb{Z}^y_2: R^{\pi}_{y}=\prod^{L}_{j=1}i\sigma^y_{j} .
\end{eqnarray}

Similar to the original LSM theorem,  the twisting operator $U_{\textbf{twist}}$ of $U(1)_z$  is given by:
\begin{eqnarray}
U_{\textbf{twist}}=\exp(\frac{2\pi i}{L}\sum^{L}_{j=1} j \bold{S}^z_{j})=\exp(\frac{\pi i}{L}\sum^{L}_{j=1} j \sigma^{z}_j).
\end{eqnarray}

Suppose the Hamiltonian \eqref{triple-product}  with length $L$ has a uniquely gapped ground state  $|\text{G.S.}\rangle$ under periodic boundary condition.  To show the excited state $U_{\textbf{twist}}|\text{G.S.}\rangle$ has low energy,  we consider the term $H_{j,j+1,j+2}$ in Hamiltonian \eqref{triple-product} involving the three neighbouring sites, $j$, $j+1$ and $j+2$.
Since $[\sigma^{z}_{j}+\sigma^{z}_{j+1}+\sigma^{z}_{j+2},H_{j,j+1,j+2}]=0$, and others commute with $H_{j,j+1,j+2}$, 
\begin{eqnarray}
&&U_{\textbf{twist}}^{+}(H_{j,j+1,j+2})U_{\textbf{twist}}\nonumber\\&&=e^{(-\pi i/L)(\sigma^{z}_{j}+2\sigma^{z}_{j+1}+3\sigma^{z}_{j+2})}H_{j,j+1,j+2} e^{(\pi i/L)(\sigma^{z}_{j}+2\sigma^{z}_{j+1}+3\sigma^{z}_{j+2})}
\nonumber\\&&=H_{j,j+1,j+2}-\frac{\pi i}{L}[\sigma^{z}_{j}+2\sigma^{z}_{j+1}+3\sigma^{z}_{j+2},H_{j,j+1,j+2}]+O(\frac{1}{L^2})\nonumber\\
&&=H_{j,j+1,j+2}-[\frac{\pi i}{L}\sum^{L}_{j=1} j \sigma^{z}_j,H_{j,j+1,j+2}]+O(\frac{1}{L^2}).\label{reminder}
\end{eqnarray}
 We can sum over all terms in \eqref{reminder}:\begin{eqnarray}
\langle\text{G.S.}|U_{\textbf{twist}}^{+}HU_{\textbf{twist}}|\text{G.S.}\rangle=E_{\text{gs}}+O(\frac{1}{L}).\label{gs}
\end{eqnarray}
The reminder $O(\frac{1}{L})$ comes from  third term in \eqref{reminder} and $\mathcal{T}_{2}$ symmetry  ensures contributions from $H_{j,j+1,j+2}$  doesn't depend on site.
Therefore we have $U_{\textbf{twist}}|\text{G.S.}\rangle=e^{i\alpha}|\text{G.S.}\rangle+O(\frac{1}{L})|\psi\rangle$.

Moreover, since $|\text{G.S.}\rangle$ preserves symmetries in \eqref{eq:sym}, it is invariant under the combination of $\mathcal{T}_{2}$ and $\mathbb{Z}^y_2$:
  \begin{eqnarray}
  T_2R^{\pi}_{y}|\text{G.S.}\rangle=TK|\text{G.S.}\rangle=|\text{G.S.}\rangle.
  \end{eqnarray}
 
Under the periodic boundary condition, we can obtain
\begin{eqnarray}
(TK)^{-1} U_{\textbf{twist}} TK&&=K^{-1}U_{\textbf{twist}}\exp(\frac{\pi i}{L}\sum^{L}_{j=1}  \sigma^{z}_j)\exp(i\pi \sigma^{z}_1)K\nonumber\\&&=-U_{\textbf{twist}}^{+}\exp(-\frac{\pi i}{L}\sum^{L}_{j=1}  \sigma^{z}_j).
\end{eqnarray}
Then 
\begin{eqnarray}
&&(TK)^{-1} U_{\textbf{twist}} TK|\text{G.S.}\rangle\nonumber\\&&=-U_{\textbf{twist}}^{+} |\text{G.S.}\rangle=-e^{-i\alpha}|\text{G.S.}\rangle-O(\frac{1}{L})|\psi'\rangle.\label{other}
\end{eqnarray}
The first equation comes from the fact $R^{\pi}_{y}U_{\theta}=U^{+}_{\theta}R^{\pi}_y$ and $U_0=\mathbb{I}$.

On the other hand, we can calculate this equation directly:
\begin{eqnarray}
(TK)^{-1} U_{\textbf{twist}} TK|\text{G.S.}\rangle&&=(TK)^{-1} (e^{i\alpha}|\text{G.S.}\rangle+O(\frac{1}{L})|\psi\rangle)\nonumber\\&&=e^{-i\alpha}|\text{G.S.}\rangle+O(\frac{1}{L})TK|\psi\rangle.\label{direct}
\end{eqnarray}
When $L$ goes to infinity, there will be a conflict between \eqref{other} and \eqref{direct}. Therefore, the energy spectrum must be either gapless or gapped with nontrivial ground state degeneracy.

		\section{Vector rotation on the $y$-direction}
		In this appendix, we will use the affine SU(2) Lie algebra to show how the primary terms transform after vector SU(2) rotation.
		For U(1) boson theory with the self-dual radius, the primary terms are just $\exp(\pm i\theta/\sqrt{2})$ and $\exp(\pm i\phi/\sqrt{2})$. For simplicity, we also represent them in the chiral language: $\exp(\pm i(\varphi_{L}+\varphi_{R})/\sqrt{2})$ and  $\exp(\pm i(\varphi_{L}-\varphi_{R})/\sqrt{2})$.
		
		The zero modes of affine SU(2) algebra can be rewritten as $J^{0}_{a}=\oint  dz J_{a}(z)$. The commutation relationship between zero modes and other fields is related to the OPEs as follows:
		\begin{eqnarray}
		[J^{0}_{a},b(w)]=\oint_{w}  dz J_{a}(z)b(w).
		\end{eqnarray}
		
		Therefore the commutation relationship between zero modes of affine SU(2) algebra and chiral primary field can be calculated using OPEs and the result is given by:
		\begin{eqnarray}
		&&[H^{0},e^{\pm i\frac{\sqrt{2}}{2}\varphi_{L}}]=\pm e^{\pm i\frac{\sqrt{2}}{2}\varphi_{L}},\quad
		\lbrack E^{0,\pm},e^{\mp i\frac{\sqrt{2}}{2}\varphi_{L}}\rbrack=2i e^{\pm i\frac{\sqrt{2}}{2}\varphi_{L}},\quad
		\lbrack E^{0,\pm},e^{\pm i\frac{\sqrt{2}}{2}\varphi_{L}}\rbrack=0, \nonumber\\
		&&\lbrack \bar{H}^{0},e^{\pm i\frac{\sqrt{2}}{2}\varphi_{R}}\rbrack=\mp e^{\pm i\frac{\sqrt{2}}{2}\varphi_{R}},\quad
		\lbrack \bar{E}^{0,\pm},e^{\pm i\frac{\sqrt{2}}{2}\varphi_{R}}\rbrack= 2ie^{\mp i\frac{\sqrt{2}}{2}\varphi_{R}},\quad
		\lbrack \bar{E}^{0,\pm},e^{\mp i\frac{\sqrt{2}}{2}\varphi_{R}}\rbrack=0.\nonumber\\
		&&~
		\end{eqnarray}
		
		Now we can perform the vector SU(2) rotation along $y$-axis whose generator is given by:
		\begin{eqnarray}
		S^{y}=\frac{i}{2}(E^{0,+}-E^{0,-}+\bar{E}^{0,+}-\bar{E}^{0,-}).
		\end{eqnarray}
		
		For $\pi/2$ rotation, the transformation can be written in the language of the generator:
		\begin{eqnarray}
		U(\frac{\pi}{2})=\frac{\sqrt{2}}{2}(1+i S^{y}).
		\end{eqnarray}
		
		For simplicity, we consider the real and imaginary parts of primary fields: $\cos({\phi}/\sqrt{2}), \sin(\phi/\sqrt{2}),\cos(\theta/\sqrt{2}), \sin(\theta/\sqrt{2})$.
		There are two primary fields invariant under rotation since they commute with $S^{y}$:
		\begin{eqnarray}
		U^{-1}\cos(\frac{\sqrt{2}}{2}\phi)U=\cos(\frac{\sqrt{2}}{2}\phi),
		U^{-1}\sin(\frac{\sqrt{2}}{2}\theta)U=\sin(\frac{\sqrt{2}}{2}\theta).
		\end{eqnarray}
		
		The other two primary fields satisfy the following commutation relationship:
		\begin{eqnarray}
		\lbrack S^{y},\cos(\frac{\sqrt{2}}{2}\theta)\rbrack=2i\sin(\frac{\sqrt{2}}{2}\phi)\ ,\lbrack S^{y},\sin(\frac{\sqrt{2}}{2}\phi)\rbrack=-2i\cos(\frac{\sqrt{2}}{2}\theta),
		\end{eqnarray}
		thus they transform under the $\pi/2$ rotation as follows:
		\begin{eqnarray}
		U^{-1}\sin(\frac{\sqrt{2}}{2}\phi)U=\cos(\frac{\sqrt{2}}{2}\theta),
		U^{-1}\cos(\frac{\sqrt{2}}{2}\theta)U=-\sin(\frac{\sqrt{2}}{2}\phi).
		\end{eqnarray}
	\end{widetext}

\end{document}